\documentclass{emulateapj}

\usepackage{graphicx}
\usepackage{amsmath}
\usepackage{amsfonts}
\usepackage{amssymb}
\usepackage{epsfig}
\usepackage{float}
\usepackage{makeidx}
\usepackage{mathrsfs}
\usepackage{natbib}
\newcommand{\begeq}{\begin{equation}}
\newcommand{\fineq}{\end{equation}}
\newcommand{\begeqarray}{\begin{eqnarray}}
\newcommand{\fineqarray}{\end{eqnarray}}
\newcommand{\xmin}{x_{\rm min}}
\newcommand{\xmax}{x_{\rm max}}

\newcommand{\green}{f_{_{\rm G}}}
\def\Green{F_{_{\rm G}}}
\def\sgreen{f_{_{\rm G}}^{\rm S}}
\def\Ngreen{N_e^{\rm S}}
\def\sigmaT{\sigma_{\rm T}}
\def\xmax{x_{\rm max}}
\def\xmin{x_{\rm min}}

\def\sgreen{f_{_{\rm G}}^{\rm S}}

\newcommand{\gapprox}{\lower.4ex\hbox{$\;\buildrel
>\over{\scriptstyle\sim}\;$}}
\newcommand{\lapprox}{\lower.4ex\hbox{$\;\buildrel
<\over{\scriptstyle\sim}\;$}}

\slugcomment{accepted by ApJ}

\shorttitle{X-Ray Time Lags in Blazars}

\shortauthors{Lewis, Becker, Finke}

\begin{document}

\title{Time-Dependent Electron Acceleration in Blazar Transients: X-ray Time Lags and Spectral Formation}

\author{Tiffany R. Lewis}

\author{Peter A. Becker}

\affil{Department of Physics and Astronomy, George Mason University, Fairfax, VA 22030-4444, USA; pbecker@gmu.edu, tlewis13@gmu.edu}

\author{Justin D. Finke}

\affil{U.S. Naval Research Laboratory, Code 7653, 4555 Overlook Avenue SW, 
Washington, DC 20375-5352, USA; justin.finke@nrl.navy.mil}

\begin{abstract}
Electromagnetic radiation from blazar jets often displays strong variability, extending from radio to $\gamma$-ray frequencies. In a few cases, this variability has been characterized using Fourier time lags, such as those detected in the X-rays from Mrk~421 using {\it Beppo}SAX. The lack of a theoretical framework to interpret the data has motivated us to develop a new model for the formation of the X-ray spectrum and the time lags in blazar jets based on a transport equation including terms describing stochastic Fermi acceleration, synchrotron losses, shock acceleration, adiabatic expansion, and spatial diffusion. We derive the exact solution for the Fourier transform of the electron distribution, and use it to compute the Fourier transform of the synchrotron radiation spectrum and the associated X-ray time lags. The same theoretical framework is also used to compute the peak flare X-ray spectrum, assuming that a steady-state electron distribution is achieved during the peak of the flare. The model parameters are constrained by comparing the theoretical predictions with the observational data for Mrk~421. The resulting integrated model yields, for the first time, a complete first-principles physical explanation for both the formation of the observed time lags and the shape of the peak flare X-ray spectrum. It also yields direct estimates of the strength of the shock and the stochastic MHD wave acceleration components in the Mrk~421 jet.
\end{abstract}

% The different journals have different requirements for keywords. The
% keywords.apj file, found on aas.org in the pubs/aastex-misc directory,
% contains a list of keywords used with the ApJ and Letters. These are
% usually assigned by the or, but authors may include them in their
% manuscripts if they wish.


\section{INTRODUCTION}

Blazars are active galactic nuclei (AGNs) possessing relativistic jets aligned with the line of sight to the observer, and emitting strongly across the entire electromagnetic spectrum. Various emission mechanisms are thought to dominate in different frequency ranges (see B\"ottcher 2007 for a review), and the spectra usually exhibit a double-peaked shape, with one peak located in the infrared to X-ray range and the other at $\gamma$-ray energies. The low-energy peak is thought to represent direct synchrotron emission from the relativistic electrons in the jet, and the high-energy emission is probably created via the Compton upscattering of the synchrotron photons, or by the upscattering of photons from the infrared through X-ray regimes, impinging on the jet from an external source, such as the surrounding accretion disk (Dermer et al. 1992; Dermer \& Schlickeiser 1993), the broad-line region (BLR; Sikora et al. 1994), or the dust torus (Kataoka et al. 1999; Blazejowski et al. 2000; Diltz \& B\"ottcher 2014).

Previous efforts to study quiescent broadband (radio to $\gamma$-ray) emission from blazar sources have focused mainly on the production of radiation via direct synchrotron emission, combined with Compton scattered emission. For example, Finke et al. (2008) employed a synchrotron/self-Compton (SSC) model to account for the optical to $\gamma$-ray emission from BL Lac objects Mrk~421 and PKS~2155-304. Using this model, they were able to deduce the energy distribution of the radiating electron population. However, the model did not attempt to account for the shape of the electron distribution using a first-principles physical approach.

There is also considerable uncertainty about the location where the observed $\gamma$-ray emission is created. If the seed photons originate in the BLR, then reverberation mapping suggests that the emission region is located $\sim 0.1\,$pc from the black hole (e.g., Bentz et al. 2006, 2013). On the other hand, the possible association between $\gamma$-ray flares and subsequent brightening of the 43\,GHz radio emission suggests an origin further out, at $\sim 1\,$pc, in which case the dust torus provides the seed photons (Nenkova et al. 2008a,b). In the SSC interpretation, the distance is not as strongly constrained (Zacharias \& Schlickeiser 2012). In high-peaked BL Lac objects such as Mrk~421, the entire spectrum is likely due to a combination of direct synchrotron and SSC emission, without any component due to the upscattering of externally produced photons, because these sources don't exhibit strong external radiation fields from the disk, the BLR (e.g. March\~a et al. 1996), or the dust torus (e.g. Plotkin et al. 2012).

Power spectral densities (PSDs) and time lags are often used to characterize the variability of blazar spectra. However, studies of the variability in the $\gamma$-ray region are restricted to timescales of a few days or
longer due to the limited sensitivity of the {\it Fermi}-LAT detector. These timescales are too long to effectively probe the region of the jet where the relativistic electrons are accelerated. Alternatively, we can probe much shorter timescales by focusing instead on the X-ray emission from blazars. For example, Zhang (2002) examined the 1998 April 21 flare of Mrk~421, observed using {\it Beppo}SAX. He utilized a Fourier-based cross-correlation function technique, and time-resolved spectral analysis, to discover hard time lags of about an hour in the X-ray emission from this source. Zhang (2002) was able to determine that the lags were not an artifact of Poisson or red noise. However, he found that Poisson noise and sparse sampling could contribute significantly to the uncertainty of the lags. Both hard and soft time lags of about an hour were also observed in the X-ray signals from the blazars PKS 2155-304 and Mrk 501 (Zhang et al. 2002; Tanihata et al. 2001; Fossati et al. 2000a).

The hard time lags of about an hour found by Zhang (2002) in observations of Mrk~421 indicate that the higher-energy X-rays are escaping from the source later than the lower-energy photons. A steady-state emission spectrum can never generate Fourier time lags, and therefore the observation of X-ray time lags in Mrk~421 necessarily implies variability in the source (e.g, Kroon \& Becker 2014, 2016). The time lags could result from the gradual upscattering of soft seed photons by a steady-state population of energetic electrons in a blob of jet plasma, but it seems more likely that they are caused by a time-dependent acceleration process, in which relativistic electrons are injected and subsequently accelerated to higher energies, radiating higher energy photons as they are accelerated (Zhang 2002). In this scenario, the observed time lags represent variability in the underlying electron distribution, and one would therefore expect to see correlated variability in the X-ray and $\gamma$-ray signals. Abdo et al. (2011) analyzed the correlated variability in the X-ray and $\gamma$-ray regimes for Mrk~421, and found no correlation, but their analysis was limited by the selection of three-day time bins due to the sensitivity of the {\it Fermi}-LAT instrument. Hence, Abdo et al. (2011) would not have been able to detect variability in the $\gamma$-ray signal on the hour-long timescales associated with the X-ray time lags, whether or not the variability was actually present.

In principle, the X-ray time lags contain detailed information about variations in the electron acceleration and the jet structure on very short timescales. However, this information cannot be utilized in the absence of a detailed quantitative model. This has motivated us to develop a new model for the evolution of the electron distribution in a blazar jet, based on a transport equation that includes terms describing second-order Fermi acceleration, synchrotron radiation, shock acceleration, adiabatic losses, and spatial diffusion.

We are specifically interested in determining whether a single physical transport model can simultaneously account for both the shape of the peak flare X-ray spectrum, and the dependence of the observed X-ray time lags on the Fourier frequency in Mrk~421. Since our goal is to develop a theoretical interpretation for the observed Fourier X-ray time lags, it is convenient for us to solve the electron transport equation in the Fourier domain. In order to render the calculation tractable, we simplify the spatial geometry by employing a one-zone model that represents an average over the radiating volume in the source, which is assumed to be a co-moving blob of plasma containing a distribution of relativistic electrons, magnetohydrodynamical (MHD) waves, and shocks (e.g., Finke et al. 2008).

The paper is organized as follows. In Section~2 we discuss the Fourier time lag concept, and the physical processes included in our model. In Section~3 we show how these processes are described within the context of the time-dependent electron transport equation, and we solve this equation to obtain the closed-form solution for the Fourier transform of the electron Green's function, which is needed to compute the X-ray time lags. We note that we do not need to compute the time-dependent electron distribution itself in order to generate the theoretical predictions for the Fourier time lags, which is a major advantage of the method employed here. In Section~4 we solve the steady-state transport equation to obtain the time-independent electron distribution and the associated Fokker-Planck coefficients. In Section~5, the resulting physical solution for the steady-state electron distribution is used to compute the associated X-ray flare spectrum. This provides a new alternative to the traditional approach which involves deducing the electron distribution by working backwards from the X-ray spectrum. We also develop the formulas required to transform the X-ray spectrum and the time lags from the co-moving frame of the outflowing plasma blob into the frame of the observer at infinity. In Section~6 we use our new model to interpret the 1998 April 21 flare from Mrk~421, based on a comparison between the theoretical predictions and the observed X-ray spectrum and time lags. We also discuss the results obtained for the various theoretical parameters. In Section~7 we reexamine the model assumptions and relate the theoretical parameters to the physical properties of the jet. Finally, in Section~8 we conclude with a summary of our main results and a discussion of our plans for future research.

\section{PHYSICAL BACKGROUND}

Recently, Finke \& Becker (2014, 2015) developed the first physical model for the generation of blazar time lags by employing a first-principles transport equation to calculate the electron energy distribution, which was then used to compute the predicted radiation time lags in the X-ray through $\gamma$-ray region. The transport equation considered by these authors included terms describing particle escape, and energy losses due to synchrotron emission, inverse-Compton scattering of external radiation, and SSC processes. In the simplest version of the model, it was assumed that the electrons are injected instantaneously with a monoenergetic distribution, but this assumption was later relaxed to treat the case of power-law injection with a random time envelope, resulting in a colored noise component. However, particle acceleration was not included in the model, and therefore it was only able to produce soft time lags, in which the injected high-energy electrons radiate a sequence of photons of diminishing energy as they cool. This behavior is consistent with some of the observations of blazars (e.g., Zhang et al. 2002), but it cannot explain the hard time lags detected by Zhang (2002) in his analysis of the X-ray data for Mrk~421.

As Zhang (2002) pointed out, the observations of hard time lags in the X-ray spectrum of Mrk~421 seem to indicate the action of time-dependent particle acceleration. In this scenario, low-energy electrons are injected into the jet, perhaps as a consequence of magnetic reconnection events (Giannios et al. 2009). These electrons are subsequently accelerated via repeated shock crossings inside the jet, and they may also experience acceleration via interactions with MHD waves, and losses due to adiabatic expansion in the jet. Diltz \& B\"ottcher (2014) studied the evolution of the electron distribution in low-frequency-peaked blazars using a time-dependent simulation that included second-order Fermi acceleration. Their model is able to produce hard time lags, but the complexity of the simulation makes it somewhat difficult to track the specific effects of the various physical processes involved. While they impose a power-law particle injection spectrum, which could simulate pre-acceleration by a shock, the Fokker-Planck equation adopted by these authors does not include a term explicitly describing first-order Fermi acceleration at an imbedded shock front. This makes it more challenging to develop a direct correspondence between their transport equation and models for jets that contain shocks (e.g., Zhang et al. 2015; Zdziarski et al. 2015; Granot \& K\"onigl 2001; K\"onigl 1981).

In order to explore the possible connection between particle acceleration and the production of hard X-ray time lags in blazars, in this paper, we extend the approach introduced by Finke \& Becker (2014, 2015) by focusing on a new, generalized transport equation that includes terms describing first-order Fermi acceleration due to shocks; second-order (stochastic) Fermi acceleration due to wave-particle interactions; particle escape; and energy losses due to synchrotron emission, inverse-Compton scattering of external radiation, and adiabatic expansion. We will also replace the constant escape timescale used by Finke \& Becker (2014, 2015) with a more physically motivated, energy-dependent timescale based on the concept of Bohm diffusion, in which the electron mean free path is essentially equal to the Larmor radius. The mathematical method is based on the development of exact analytical solutions to the linear transport equation, and therefore SSC losses are not included, since they are inherently nonlinear.

The X-ray time lags observed from Mrk~421 and reported by Zhang (2002) reveal the presence of transients with a variability timescale of about one hour, corresponding to a $\sim 10\%$ change in amplitude (Fossati et al. 2000b). The overall shape of the X-ray spectrum varies on much longer timescales of about one day. This suggests that while the time lags require consideration of impulsive particle injection, we may be able to model the peak flare X-ray spectrum using a steady-state model in which the electrons are continuously injected. Hence we will treat the electron distribution in the radiating plasma blob as the sum of a variable component and a steady-state component. Since our transport equation includes synchrotron losses, the synchrotron X-ray spectra computed using our solution for the electron distribution are self-consistent.

\subsection{Fourier Time Lags}

The X-ray time lags from Mrk~421 discovered by Zhang (2002) were computed using the Fourier-based technique first pioneered by van der Klis et al. (1987) in his study of black-hole variability. The method requires the evaluation of the complex cross spectrum, $C$, which is defined by 
\begin{equation}
C(\epsilon_s,\epsilon_h,\omega) = G_s^*(\omega) G_h(\omega) \ ,
\label{eq1}
\end{equation}
where $\omega$ denotes the Fourier frequency, $G_s(\omega)$ and $G_h(\omega)$ represent the Fourier transforms of the soft- and hard-energy channel time series data, corresponding to photon energies $\epsilon_s$ and $\epsilon_h$, respectively, and the asterisk represents the complex conjugate. The phase angle of the complex cross spectrum is given by
\begin{equation}
\phi = {\rm Arg}(C) \ ,
\label{eq2}
\end{equation}
where the argument of the complex variable $z=x+i y$ is defined by the relation
\begin{equation}
{\rm Arg}(x+iy) \equiv \tan^{-1}\left(\frac{y}{x}\right) \ .
\label{eq3}
\end{equation}
The associated Fourier time lag, $\delta t$, is computed using
\begin{equation}
\delta t = \frac{\phi}{\omega} \ .
\label{eq4}
\end{equation}

It is straightforward to show that if the hard channel time series has the same shape as the soft channel time series, but with a delay equal to $\Delta t$, then the Fourier time lag computed using Equation~(\ref{eq4}) is $\delta t=\Delta t$, as expected (Kroon \& Becker 2014). Furthermore, the detection of a finite time lag $\delta t$ implies the presence of actual variability in the X-ray signal, since without variability, the Fourier time lag formally reduces to $\delta t=0$ (Kroon \& Becker 2016). In the application of interest here, the variability is associated with a transient flare produced in the blazar jet. During the transient, relativistic electrons are impulsively injected into a blob of jet plasma, as a result of magnetic reconnection or some other instability. The injected particles are subject to acceleration and radiative losses until they escape from the blob, and the time-dependent nature of this process gives rise to the observed time lags.

Our primary goal in this paper is to compute the time lags as a function of the Fourier frequency so that they can be compared with the data analyzed by Zhang (2002) during the 1998 April 21 X-ray flare of Mrk~421. A secondary goal of the paper is to use the same transport equation to compute the steady-state X-ray spectrum emitted by a population of electrons that is continually injected into the jet plasma, which will be compared with the peak X-ray spectrum observed during the same X-ray flare from which the time lag data was derived. Analysis of the two resulting sets of theoretical parameters should yield insight into the nature of the physical processes occurring in the plasma during the observed transient.

\subsection{Spatial Diffusion}

In the scenario envisioned here, two different types of interactions control the energetic and spatial aspects of the stochastic particle transport. On small spatial scales, the particles interact with MHD waves propagating in the local magnetic field, which results in second-order Fermi acceleration, and also regulates the spatial transport. The mean-free path on small scales is therefore equal to the coherence length for the MHD turbulence, denoted by $\ell_{_{\rm MHD}}$. On large spatial scales, the particle transport occurs via diffusion, with a mean-free path dictated by the relativistic electron's Larmor radius, $r_{\rm L}$, defined by
\begin{equation}
r_{\rm L} \equiv \frac{E}{q B} \ ,
\label{eq5}
\end{equation}
where $E$ is the electron energy, $q$ is the magnitude of the electron charge, and $B$ is the magnetic field inside the plasma blob. This physical regime corresponds to Bohm diffusion. Once the electron's energy gets sufficiently large, the Larmor radius becomes comparable to the radius of the plasma blob, $R$, and the particles are able to escape. Hence no further acceleration occurs once $r_{\rm L} \sim R$, which is a statement of the Hillas (1984) condition.

In the simple one-zone model considered here, the timescale for the electrons to escape from the acceleration region via Bohm diffusion, denoted by $t_{\rm esc}$, is given by
\begin{equation}
t_{\rm esc} = \frac{R}{w_{\rm L}} \ ,
\label{eq6}
\end{equation}
where the Bohm diffusion velocity, $w_{\rm L}$, for the relativistic electrons is defined by
\begin{equation}
w_{\rm L} = \frac{c}{R/r_{\rm L}} \ ,
\label{eq7}
\end{equation}
and $r_{\rm L}$ is given by Equation~(\ref{eq5}). We note that the Hillas condition, $r_{\rm L} < R$, automatically limits the diffusion velocity so that $w_{\rm L} < c$ in the blob's frame, as required to maintain causality. We will revisit this constraint in Section~7.4. By combining Equations~(\ref{eq5}), (\ref{eq6}), and (\ref{eq7}), we obtain for the escape timescale
\begin{equation}
t_{\rm esc} = \frac{R^2}{c \, r_{\rm L}} = \frac{R^2 q B}{c E} \ ,
\label{eq8}
\end{equation}
which indicates that the high-energy particles escape preferentially, since they have the largest Larmor radii. In the ultrarelativistic case, $t_{\rm esc}$ can be evaluated as a function of the particle momentum $p=E/c$ using
\begin{equation}
t_{\rm esc}(p) = \frac{R^2 q B}{c^2 p} \ .
\label{eq8b}
\end{equation}
\subsection{MHD Acceleration}

The electrons experience both first-order Fermi acceleration (if a shock is present), and also second-order (stochastic) Fermi acceleration if they interact with a random field of magnetic irregularities. In either case, the scattering centers are MHD waves, and therefore we must ensure that the acceleration rate implied by our model does not lead to particle energies exceeding the radiation reaction limit (e.g., Cerutti et al. 2012). MHD wave acceleration cannot boost particles beyond this energy, although it should be noted that electrostatic acceleration is not bound by this constraint (Kroon et al. 2016; Cerutti et al. 2012).

The maximum particle energy consistent with the radiation reaction constraint is determined by setting the Larmor gyroperiod, which is the minimum timescale for MHD wave acceleration, equal to the synchrotron loss timescale. The Larmor gyration timescale, $t_{\rm L}$, for the relativistic electrons is defined by
\begin{equation}
t_{\rm L} \equiv \frac{2\pi r_{\rm L}}{c} \ .
\label{eq9}
\end{equation}
The synchrotron timescale, $t_{\rm syn}$, is computed by considering the mean energy loss rate per electron due to synchrotron emission. The result obtained for an isotropic, ultrarelativistic electron distribution is (e.g., Rybicki and Lightman 1979)
\begin{equation}
<\!\dot{E}\!>_{\rm syn} = - \frac{4}{3}\frac{\sigmaT
c \, U_B}{m_e^2 c^4} E^2 \ ,
\label{eq10}
\end{equation}
where $\sigmaT = (8\pi /3) q^4/m_e^2 c^4$ is the Thomson cross section and $U_B = B^2/8\pi$ is the magnetic energy density. The characteristic timescale for synchrotron losses is therefore
\begin{equation}
t_{\rm syn} = - \frac{E}{<\!\dot{E}\!>_{\rm syn}}
= \frac{3}{4} \frac{m_e^2 c^4}{\sigmaT c \,U_B} \frac{1}{E} \ .
\label{eq11}
\end{equation}

By combining Equations~(\ref{eq5}), (\ref{eq9}), and (\ref{eq11}), we find that the ratio of the Larmor and synchrotron timescales for the ultrarelativistic electrons can be written as
\begin{equation}
\frac{t_{\rm L}}{t_{\rm syn}} = \frac{\sigmaT
B E^2} {3 m_e^2 c^4 q } = 4.61 \times 10^{-6}\left( \frac{\gamma}{10^5}
\right)^{2} \left( \frac{B}{1 \, \rm G}\right) \ .
\label{eq12}
\end{equation}
In the flares observed from Mrk~421, the X-ray emission is produced by electrons with Lorentz factor $\gamma=E/m_e c^2 \sim 10^5$, radiating in a magnetic field of strength $B \sim 0.08\,$G (e.g., Abdo et al. 2011). It follows that in the energy regime relevant for these electrons, the synchrotron timescale is much longer than the Larmor timescale, i.e.,
\begin{equation}
t_{\rm syn} \gg t_{\rm L} \ .
\label{eq13}
\end{equation}
Hence during the observed X-ray flares from Mrk~421, the electron acceleration is not limited by the synchrotron radiation reaction. However, it is interesting to note that this limit does come into play when considering the extreme electron acceleration that occurs during the high-energy $\gamma$-ray flares recently observed from the Crab nebula (e.g., Kroon et al. 2016).

As discussed in Section 2.2, on small scales, the spatial diffusion of the electrons is regulated by interactions with MHD waves, with coherence length $\ell_{_{\rm MHD}}$. The associated spatial diffusion coefficient is therefore given by (Dr\"oge \& Schlickeiser 1986; Reif 1969)
\begin{equation}
\kappa = \frac{c\, \ell_{_{\rm MHD}}}{3} \ ,
\label{eq14}
\end{equation}
which is related to the momentum diffusion coefficient, $D(p)$, via (Dr\"oge et al. 1987; Schlickeiser 1985)
\begin{equation}
D(p)\kappa(p) = \frac{p^2 v_{\rm A}^2}{9} \ ,
\label{eq15}
\end{equation}
with $v_{\rm A}$ denoting the Alfv\'en velocity. Since the coherence length $\ell_{_{\rm MHD}}$ is independent of the particle momentum $p$, we can combine Equations (\ref{eq14}) and (\ref{eq15}) to show that the momentum dependence of $D$ is given by the hard-sphere relation (e.g., Park \& Petrosian 1995)
\begin{equation}
D(p) = D_0 \, p^2 \ ,
\label{eq16}
\end{equation}
where the momentum-diffusion constant $D_0$ is defined by
\begin{equation}
D_0 \equiv \frac{c}{3\, \ell_{_{\rm MHD}}} \left(\frac{v_{\rm A}}{c} \right)^2
\ \propto \ {\rm s}^{-1} \ .
\label{eq17}
\end{equation}

\subsection{First-order Fermi Processes}

In addition to the stochastic acceleration that the electrons experience as a result of interactions with a random field of Alfv\`en waves, the electrons in the plasma blob may also experience first-order Fermi acceleration due to repeated interactions with shock waves propagating along the jet axis (Achterberg et al. 2001). The particles will also experience first-order losses due to adiabatic expansion in the jet (Marscher \& Gear 1985).

\subsubsection{Shock Acceleration}

In the model envisioned here, we adopt the picture discussed by Zdziarski et al. (2015) in which shocks propagate along the axis of the jet with velocity $w$ as seen in the frame of the central galaxy, which we call the lab frame. The jet itself propagates with velocity $v = \beta c$ in the lab frame. If $w$ and $v$ are both close to the speed of light, and the shock and the jet have nearly the same Lorentz factor as measured in the lab frame, then in the frame of the shock, the upstream and downstream flows are nonrelativistic, and therefore we can use a classical prescription to describe the acceleration of the electrons at the shock. The shock compression ratio is defined by 
\begin{equation}
\chi \equiv \frac{u_-}{u_+} \ ,
\label{eq18}
\end{equation}
where $u_-$ and $u_+$ are, respectively, the upstream and downstream velocities measured in the frame of the shock.

The mean rate of change of the particle momentum due to shock crossings is given by (Berezhko \& Ellison 1999; Dr\"oge et al. 1987; Drury 1983; Webb et al. 1984)
\begin{equation}
<\!\dot{p}\!>_{\rm sh} = \frac{4}{3}\frac{\Delta u}{c} \frac{p}{t_{\rm cyc}} \ ,
\label{eq19}
\end{equation}
where
\begin{equation}
t_{\rm cyc} = \frac{4}{c} \left( \frac{\kappa_-}{u_-} + \frac{\kappa_+}{u_+} \right)
\label{eq20}
\end{equation}
denotes the timescale for particles to cycle across the shock, $\kappa$ is the spatial diffusion coefficient, and the velocity jump at the shock is given by
\begin{equation}
\Delta u \equiv u_- - u_+ = u_-\left( \frac{\chi-1}{\chi} \right) \ .
\label{eq21}
\end{equation}
The subscripts ``-" and ``+" designate quantities measured on the immediate upstream and downstream sides of the shock, respectively. We will assume for simplicity that the spatial diffusion coefficient remains constant across the shock, so that $\kappa_-=\kappa_+=\kappa$, although a jump in $\kappa$ can easily be incorporated.

Combining Equations (\ref{eq14}), (\ref{eq18}), (\ref{eq19}), (\ref{eq20}), and (\ref{eq21}), we arrive at
\begin{equation}
<\!\dot{p}\!>_{\rm sh} = \frac{u_-^2}{c\,\chi\ell_{_{\rm MHD}}}\left( \frac{\chi - 1}{\chi + 1} \right) \, p \ .
\label{eq22}
\end{equation}
Hence can write the mean particle acceleration rate due to shock crossings as
\begin{equation}
<\!\dot{p}\!>_{\rm sh} = A^{\rm sh}_0 \, p \ ,
\label{eq23}
\end{equation}
where the constant $A^{\rm sh}_0$ is defined by
\begin{equation}
A^{\rm sh}_0 \equiv \frac{u_-^2}{c\,\chi\ell_{_{\rm MHD}}}\left( \frac{\chi - 1}{\chi + 1} \right)
\ \propto \ \rm s^{-1} \ .
\label{eq24}
\end{equation}

\subsubsection{Adiabatic Losses}

The mean first-order momentum loss rate for electrons with momentum $p$ due to adiabatic expansion of the outflowing plasma blob is given by (Jokipii 1971; Gleeson \& Webb 1978; Becker 1992; Gupta et al. 2006)
\begin{equation}
\left<\!\frac{dp}{dt}\!\right>\bigg|_{\rm ad} = - \frac{1}{3} \Big(\vec{\nabla} \cdot \vec{v}\Big) \, p
= - \frac{1}{3V} \frac{dV}{dt} \, p \ ,
\label{eq25}
\end{equation}
where $\vec{v}$ denotes the vector velocity field of the jet, $V$ is the volume of the plasma blob, and $d/dt$ represents the co-moving time derivative. Hence the mean adiabatic loss rate can be written as
\begin{equation}
<\!\dot{p}\!>_{\rm ad} = A^{\rm ad}_0 \, p \ ,
\label{eq26}
\end{equation}
where the quantity $A^{\rm ad}_0$ is defined by 
\begin{equation}
A^{\rm ad}_0 \equiv - \frac{1}{3V} \frac{dV}{dt} \ \propto \ \rm s^{-1} \ .
\label{eq27}
\end{equation}
We will assume that $A^{\rm ad}_0$ can be treated as a constant within the relatively small volume of the radiating blob during the X-ray flare.

\section{PARTICLE TRANSPORT MODEL}

The particle transport equation we will focus on here includes terms describing stochastic acceleration, shock acceleration, particle escape, and losses due to synchrotron emission, inverse-Compton scattering of external radiation, and adiabatic expansion. The various terms were discussed in detail in Section~2. Our goal is to determine whether a physical model incorporating these particle transport processes can simultaneously explain both the production of the hard time lags and the peak X-ray spectrum observed during the 1998 April 21 flare from Mrk~421.

The solution we will obtain for the electron Fourier transform represents the time-dependent Green's function response to the {\it impulsive} injection of monoenergetic electrons into the plasma blob, possibly as a result of magnetic reconnection taking place near a shock imbedded in the plasma (Nalewajko et al. 2011; Sironi et al. 2015). Once we have obtained the exact solution for the electron Fourier transform, we will use it to compute the Fourier transform of the observed X-ray emission, under the assumption that the electrons emit synchrotron radiation. The Fourier transform of the X-ray emission is then used to compute the associated time lags.

We will also obtain the exact solution for the steady-state electron Green's function resulting from the {\it continual} injection of monoenergetic electrons into the blob, possibly picked up from the tail of the thermal electron distribution. The steady-state electron Green's function will be used to calculate the associated time-independent synchrotron X-ray spectrum, which we interpret as the peak X-ray spectrum observed during a flare, when the electrons have reached an approximate equilibrium between acceleration and energy losses. The model parameters will be constrained by comparing the computed X-ray time lags and the X-ray spectrum with the data for the 1998 April 21 flare from Mrk~421 obtained using {\it Beppo}SAX. The resulting parameter study based on the new particle transport model developed here may provide the best glimpse yet into the nature of the microphysical processes occurring in the outflowing jet plasma.

The observation of $\sim 10\%$ variability on timescales of $\sim 1\,$hour, combined with more significant changes in the shape of the X-ray spectrum occurring on longer timescales of $\sim 1\,$day, suggests the possibility of treating the electron distribution using two components (Fossati et al. 2000b). In this interpretation, a time-dependent electron component creates the $\sim 10\%$ amplitude variability on $\sim 1\,$hour timescales, and a steady-state electron component produces the remaining $\sim 90\%$ of the spectrum, with a variability timescale of $\sim 1\,$day. This approach is supported by estimates carried out in Section~7.1, where we show that the equilibration timescale for the electrons is $\sim 5-10\,$hours. This suggests that on timescales of $\sim 1\,$hour, comparable to the observed time lags, the electrons are out of equilibrium. On the other hand, the spectral component with $\sim 1\,$day variability is probably produced by electrons with a steady-state distribution. In our model, the lower-amplitude, time-dependent component represents the variable distribution resulting from impulsive electron injection, perhaps related to sporadic magnetic reconnection events occurring in the vicinity of a shock (Giannios 2013). Conversely, the steady-state population results from the continual injection of seed electrons, possibly picked up from the high-energy tail of the thermal electron distribution in the blob.

\subsection{Time-dependent Transport Equation}

The fundamental time-dependent transport equation governing the momentum distribution function, $f(p,t)$, for the relativistic electrons in the jet plasma is written in the co-moving frame as (e.g., Becker 1992; Park \& Petrosian 1995; Schlickeiser 1985)
\begin{align}
\frac{\partial f}{\partial t} =& -\frac{1}{p^2}\frac{\partial}{\partial p}
\bigg\{p^2 \bigg[-D(p)\frac{\partial f}{\partial p} + <\!\dot{p}\!>_{\rm gain}
f  \nonumber \\
& + <\!\dot{p}\!>_{\rm loss}f \bigg ] \bigg \} - \frac{f}{t_{\rm esc}(p)}
+ \dot f_{\rm source} \ ,
\label{eq28}
\end{align}
where $p$ is the electron momentum and the terms on the right-hand side describe the effects of momentum diffusion (stochastic acceleration), systematic momentum gains, systematic momentum losses, particle escape, and particle injection, respectively. The distribution function, $f(p,t)$, is related to the total number of electrons in the blob, ${\cal N}_e(t)$, via the integral
\begin{equation}
{\cal N}_e(t) = \int_{0}^{\infty} 4 \pi \, p^2 f(p,t) \, dp \ .
\label{eq29}
\end{equation}
This relation establishes the normalization of the distribution function $f$. We discuss the specific forms adopted for the various terms on the right-hand side of the transport equation below.

The first term on the right-hand side of the transport equation describes the second-order acceleration resulting from stochastic interactions between the electrons and the local MHD wave population. The process is described by the momentum diffusion coefficient $D(p)$, which is given by the hard-sphere formulation (see Equations~(\ref{eq16}) and (\ref{eq17})). The second term on the right-hand side of the transport equation describes the combined effect of the two first-order Fermi processes included in our model (adiabatic losses and shock acceleration), which are consolidated by writing
\begin{equation}
<\dot{p}>_{\rm gain} = <\dot{p}>_{\rm sh} + <\dot{p}>_{\rm ad} = A_0\,p\ ,
\label{eq31}
\end{equation}
where the constant $A_0$ is defined by (see Equations~(\ref{eq24}) and (\ref{eq27}))
\begin{equation}
A_0 \equiv A^{\rm sh}_0 + A^{\rm ad}_0
= \frac{u_-^2}{c\,\chi\ell_{_{\rm MHD}}}\left( \frac{\chi - 1}{\chi + 1} \right)
- \frac{1}{3} \, (\vec{\nabla} \cdot \vec{v}) \ .
\label{eq32}
\end{equation}

The third term on the right-hand side of the transport equation models the momentum losses experienced by the electrons due to the emission of synchrotron radiation, with a quadratic energy dependence given by Equation~(\ref{eq10}). In the blazar application treated here, the electrons are ultrarelativistic, so that $E=p c$, and therefore the associated momentum loss rate for synchrotron emission is given by
\begin{equation}
<\!\dot{p}\!>_{\rm loss} = \frac{1}{c}<\!\dot{E}\!>_{\rm loss} = - \frac{B_0}{m_ec} \, p^2 \ ,
\label{eq33}
\end{equation}
where the positive constant $B_0 \propto \rm s^{-1}$ is defined by
\begin{equation}
B_0 \equiv \frac{4}{3} \frac{\sigmaT}{m_e c} \frac{B^2}{8 \pi} \ .
\label{eq34}
\end{equation}
We note that inverse-Compton losses due to the up-scattering of external seed photons can also be included in our model if we replace $B^2/(8 \pi)$ in Equation~(\ref{eq34}) with $B^2/(8 \pi) + U_{\rm ph}$, where $U_{\rm ph}$ represents the energy density in the incident (external) photons (Rybicki \& Lightman 1979). Losses due to the upscattering of cosmic microwave background photons are completely insignificant compared with synchrotron losses in the typical blazar magnetic field $B \sim 0.01-0.1\,$G, but losses due to the upscattering of incident photons from the broad-line region or the accretion disk may be significant (e.g., Dermer et al. 1992; Dermer \& Schlickeiser 1993; Sikora et al. 1994). We also note that Equation~(\ref{eq33}) neglects SSC losses, which cannot be modeled using a linear transport equation (e.g., Finke et al. 2008). In our application to Mrk~421, we will focus on losses due to synchrotron emission only, although the effect of inverse-Compton scattering can easily be incorporated by adopting a non-zero value for $U_{\rm ph}$.

The fourth term on the right-hand side of the transport equation represents the escape of particles with mean escape timescale $t_{\rm esc}$, which is given as a function of the electron momentum $p$ by Equation~(\ref{eq8b}). The fifth term on the right-hand side of the transport equation represents the instantaneous injection of $N_0$ electrons with momentum $p_0$ into the blob at time $t_0$. The form of the source term is therefore given by
\begin{equation}
\dot f_{\rm source} = \frac{N_0 \, \delta(p-p_0)\delta(t-t_0)}{4 \pi p_0^2}
\ .
\label{eq36}
\end{equation}

Using Equations~(\ref{eq8b}), (\ref{eq16}), (\ref{eq31}), (\ref{eq33}), and (\ref{eq36}) to substitute into the transport equation~(\ref{eq28}) yields the specific time-dependent equation of interest here,
\begin{align}
\frac{\partial \green}{\partial t} =& - \frac{1}{p^2} \frac{\partial}
{\partial p} \left[p^2 \left(-D_0 p^2\frac{\partial \green}{\partial p}
+ A_0 p \green - \frac{B_0 p^2}{m_e c} \green \right) \right] \nonumber \\
&- \frac{c^2 p \green}{R^2 q B}
+ \frac{N_0 \delta(p-p_0)\delta(t-t_0)}{4 \pi p_0^2}
\ .
\label{eq37}
\end{align}
The solution to this equation is the {\it Green's function}, $\green(p,t)$, which represents the electron distribution resulting from the instantaneous injection of monoenergetic particles with momentum $p=p_0$ at time $t=t_0$.
Since the transport equation is linear, it follows that the particular solution for the electron distribution, $f$, resulting from any source distribution in time and energy, $\dot f_{\rm source}$, can be obtained via integral convolution.

In seeking an analytical solution to Equation~(\ref{eq37}), it is convenient to work in terms of the dimensionless momentum, $x$, defined by
\begin{equation}
x \equiv \frac{p}{m_e c} \ ,
\qquad
x_0 \equiv \frac{p_0}{m_e c} \ ,
\label{eq38}
\end{equation}
where $x=\sqrt{\gamma^2-1}$, and $\gamma=E/(m_e c^2)$ denotes the electron Lorentz factor. Note that for the highly relativistic ($\gamma \gg 1$) electrons of interest here, $x$ is equivalent to the Lorentz factor $\gamma$, and we will therefore use these two notations interchangeably. Transforming from $(p,t)$ to $(x,t)$ in the transport equation~(\ref{eq37}) yields
\begin{align}
\frac{1}{D_0} \frac{\partial \green}{\partial t} =& \frac{1}{x^2} \frac{\partial}
{\partial x} \left[x^2 \left(x^2\frac{\partial \green}{\partial x}
- a x \green + b x^2 \green \right) \right] - \frac{x \green}{\tau} \nonumber \\
&+ \frac{N_0 \, \delta(x-x_0) \delta(t-t_0)}{4 \pi D_0 (m_e c)^3 x_0^2} \ ,
\label{eq39}
\end{align}
where we have also introduced the new dimensionless constants $a$, $b$, and $\tau$, defined by
\begin{equation}
a \equiv \frac{A_0}{D_0} \ , \qquad
b \equiv \frac{B_0}{D_0} \ , \qquad
\tau \equiv \frac{R^2 q B D_0}{m_e c^3} \ .
\label{eq40}
\end{equation}
\subsection{Electron Fourier Transform}

Computation of the radiation time lags in the X-ray regime using Equation~(\ref{eq4}) requires the development of expressions for the Fourier transforms of the electron and photon distributions. We begin by defining the Fourier transform $\Green$ of the electron Green's function $\green$ with respect to the time $t$ using the integral expressions
\begin{equation}
\Green(x,\omega) \equiv \int_{-\infty}^{\infty} e^{i \omega t} \green(x,t) dt \ ,
\label{eq41}
\end{equation}
and 
\begin{equation}
\green(x,t) \equiv \frac{1}{2\pi}\int_{-\infty}^{\infty} e^{-i \omega t} \Green(x,\omega) d\omega \ ,
\label{eq42}
\end{equation}
where $\omega$ denotes the circular Fourier frequency, which is related to the Fourier frequency in Hertz, $\nu_f$, via
\begin{equation}
\omega \equiv 2 \pi \nu_f \ .
\label{eq43}
\end{equation}

By applying the operator $\int_{-\infty}^{\infty} e^{i \omega t} dt$ to Equation~(\ref{eq39}), we find that the Fourier transform $\Green$ is governed by the equation
\begin{align}
- \frac{i \omega}{D_0} \Green =& \frac{1}{x^2} \frac{d}{dx}\left(x^4
\frac{d \Green}{dx} - a x^3 \Green + b x^4 \Green \right)
-\frac{x}{\tau} \Green \nonumber \\
&+ \frac{N_0 \delta(x-x_0) e^{i \omega t_0}}{4 \pi D_0 (m_e c)^3 x_0^2} \ .
\label{eq44}
\end{align}
In the special case $x \ne x_0$, the source term in Equation~(\ref{eq44}) is not active, and we obtain a homogeneous, linear, second-order ordinary differential equation. The fundamental solutions to the homogeneous equation satisfying suitable boundary conditions at large and small values of $x$ are given by
\begin{equation}
\Green(x,\omega) \propto e^{-bx/2 } x^{-2 + a/2}
\begin{cases}
M_{\kappa,\mu}(bx) \ , & x \le x_0 \ , \\
W_{\kappa,\mu}(bx) \ , & x \ge x_0 \ ,
\end{cases}
\label{eq45}
\end{equation}
where $M_{\kappa,\mu}$ and $W_{\kappa,\mu}$ denote Whittaker's functions, and the constants $\kappa$ and $\mu$ are defined by
\begin{equation}
\kappa = 2-\frac{1}{b\tau} + \frac{a}{2} \ , \qquad
\mu = \sqrt{\frac{(a+3)^2}{4} - \frac{i \omega}{D_0}} \ .
\label{eq46}
\end{equation}

Next we return to consideration of the inhomogeneous version of Equation~(\ref{eq44}), with the source term included. The global solution for $\Green$ satisfying the inhomogeneous equation must be continuous across the injection energy $x=x_0$, therefore it is convenient to write the solution in the form
\begin{equation}
\Green(x) = C_0 \, e^{-b x/2 } x^{-2 + a/2}
M_{\kappa,\mu}(b \xmin) \,
W_{\kappa,\mu} (b \xmax) \ ,
\label{eq47}
\end{equation}
where
\begin{equation}
\xmin \equiv \min(x,x_0) \ , \qquad
\xmax \equiv \max(x,x_0) \ ,
\label{eq48}
\end{equation}
and the normalization constant $C_0$ is determined by applying the derivative jump condition implied by the source term in the transport equation~(\ref{eq44}). Integration of the transport equation with respect to $x$ over a small range surrounding the injection momentum $x_0$ yields the derivative jump condition
\begin{equation}
\lim_{\delta \to 0} \left[ \frac{dF}{dx}\right]_{x_0+\delta}
- \lim_{\delta \to 0} \left[ \frac{dF}{dx}\right]_{x_0-\delta}
= - \frac{N_0 e^{i \omega t_0}}{4 \pi D_0 (m_e c)^3 x_0^4} \ .
\label{eq49}
\end{equation}
Substituting Equation~(\ref{eq47}) into Equation~(\ref{eq49}) yields
\begin{align}
C_0 &\, e^{-b x_0/2 } x_0^{-2 + a/2} \, b \,
\big[M_{\kappa,\mu}(bx_0) W'_{\kappa,\mu}(bx_0) \nonumber \\
&- W_{\kappa,\mu}(bx_0) M'_{\kappa,\mu}(bx_0)\big]
= - \frac{N_0 e^{i \omega t_0}}{4 \pi D_0 (m_e c)^3 x_0^4} \ .
\label{eq50}
\end{align}

The Wronskian appearing inside the square brackets on the right-hand side of Equation~(\ref{eq50}) can be evaluated using the identity (Abramowitz \& Stegun 1970; Slater 1960)
\begin{equation}
M_{\kappa,\mu}(z)W'_{\kappa,\mu}(z)
- W_{\kappa,\mu}(z)M'_{\kappa,\mu}(z)
= - \dfrac{\Gamma(1+2\mu)}{\Gamma(\mu-\kappa+1/2)} \ .
\label{eq51}
\end{equation}
Combining relations, the solution obtained for the normalization constant $C_0$ is
\begin{equation}
C_0 = \frac{N_0 e^{i \omega t_0} e^{b x_0/2}}
{4 \pi b D_0 (m_e c)^3 x_0^{2+a/2}} \frac{\Gamma(\mu-\kappa+1/2)}{\Gamma(1+2\mu)} \ .
\label{eq52}
\end{equation}
Using this result to substitute for $C_0$ in Equation~(\ref{eq47}) yields the final solution for the electron Fourier transform,
\begin{align}
\Green(x) =& \frac{N_0 e^{i \omega t_0} e^{b x_0/2}}
{4 \pi b D_0 (m_e c)^3 x_0^{2+a/2}} \frac{\Gamma(\mu-\kappa+1/2)}{\Gamma(1+2\mu)}
\, e^{-b x/2 } \nonumber \\
&\times x^{-2 + a/2}
M_{\kappa,\mu}(b \xmin) \,
W_{\kappa,\mu} (b \xmax) \ ,
\label{eq53}
\end{align}
where $\kappa$ and $\mu$ are evaluated using Equations~(\ref{eq46}) and $\xmin$ and $\xmax$ are defined by Equations~(\ref{eq48}). Equation~(\ref{eq53}) gives the electron Fourier transform in the co-moving frame of the outflowing plasma blob.

In our astrophysical applications, it is convenient to work in terms of the electron number distribution, $N_e$, which is related to the distribution function $\green$ via
\begin{equation}
N_e(x,t) = 4 \pi \, (m_e c)^3 x^2 \green(x,t) \ .
\label{eq54}
\end{equation}
The corresponding total number of electrons in the blob at time $t$, denoted by ${\cal N}_e(t)$, can be computed from $N_e(x,t)$ using (cf. Equation~(\ref{eq29}))
\begin{equation}
{\cal N}_e(t) = \int_0^\infty N_e(x,t) \, dx \ .
\label{eq55}
\end{equation}
The Fourier transform of the electron number distribution with respect to time $t$ is defined by
\begin{equation}
\tilde N_e(x,\omega) \equiv \int_{-\infty}^\infty e^{i \omega t} \, N_e(x,t) \, dt
= 4 \pi \, (m_e c)^3 x^2 \Green(x,\omega) \ ,
\label{eq56}
\end{equation}
where the final result follows from Equations~(\ref{eq41}) and (\ref{eq54}). Using Equation~(\ref{eq53}) to substitute for $\Green$ in Equation~(\ref{eq56}) yields the exact solution for the Fourier transform of the electron number distribution, given by
\begin{align}
\tilde N_e(x,\omega) =& \frac{N_0 e^{i \omega t_0} e^{b x_0/2}}
{b D_0 x_0^{2+a/2}} \frac{\Gamma(\mu-\kappa+1/2)}{\Gamma(1+2\mu)}
\, e^{-b x/2 } x^{a/2} \nonumber \\
& \times M_{\kappa,\mu}(b \xmin) \,
W_{\kappa,\mu} (b \xmax)
\ .
\label{eq57}
\end{align}
Equation~(\ref{eq57}) gives the exact solution for the Fourier transform, $\tilde N_e(x,\omega)$, of the time-dependent electron number distribution, $N_e(x,t)$, resulting from the impulsive injection of monoenergetic particles into the plasma blob at time $t_0$. We will derive the corresponding expression for the Fourier transform of the radiated synchrotron spectrum in Section~5, and the resulting X-ray time lags will be computed in Section~6 and plotted in Figure~1.

\section{STEADY-STATE ELECTRON DISTRIBUTION}

The X-ray time lags from Mrk~421 reported by Zhang (2002) are produced by transients with a variability timescale of about one hour. Examination of Figure~3 in Fossati et al. (2000b), which gives the X-ray spectrum during the same time interval analyzed by Zhang (2002), indicates that the associated variation of the spectrum on $\sim 1\,$hour timescales is about 10\% of the maximum flux. The observations show that the overall shape of the X-ray spectrum changes significantly on much longer timescales of $\sim 1\,$day. This suggests that while the time lags require consideration of impulsive particle injection, we may be able to simulate the peak flare X-ray spectrum using a steady-state model in which the electrons are continuously injected. In this section, we compute the time-independent synchrotron spectrum produced by a steady-state distribution of relativistic electrons accelerated in the jet. It is important to emphasize that our steady-state hypothesis implicitly assumes that the continually-injected electrons reach equilibrium at the peak of the flare, as further discussed in Section~7.1.

\subsection{Time-independent Transport Equation}

The transport equation satisfied by the steady-state Green's function, $\sgreen(p)$, resulting from the continual injection of $\dot N_0$ particles per second with momentum $p=p_0$ can be written as (cf. Equation~(\ref{eq37}))
\begin{align}
\frac{\partial \sgreen}{\partial t} =& 0 = - \frac{1}{p^2} \frac{d}{d p} \bigg[p^2 \bigg(-D_0 p^2\frac{d \sgreen}{d p}
+ A_0 p \sgreen \nonumber \\
&- \frac{B_0}{m_e c} p^2 \sgreen \bigg) \bigg]
- \frac{c^2 p \sgreen}{R^2 q B}
+ \frac{\dot N_0 \delta(p-p_0)}{4 \pi p_0^2}
\ .
\label{eq58}
\end{align}
Transforming from $p$ to the dimensionless momentum, $x$, where (see Equations~(\ref{eq38}))
\begin{equation}
x \equiv \frac{p}{m_e c} \ , \qquad x_0 \equiv \frac{p_0}{m_e c} \ ,
\label{eq59}
\end{equation}
yields
\begin{align}
\frac{\partial \sgreen}{\partial t}
=& 0 = - \frac{1}{x^2} \frac{d}{d x} \bigg[x^2 \bigg(-D_0 x^2
\frac{d \sgreen}{d x} + A_0 x \sgreen \nonumber \\
&- B_0 x^2 \sgreen \bigg) \bigg] - \frac{m_e c^3 x \sgreen}{R^2 q B}
+ \frac{\dot N_0 \delta (x-x_0)}{4 \pi (m_e c)^3 x_0^2} \ .
\label{eq60}
\end{align}
Proceeding as in Section~3, we introduce the dimensionless constants $a$, $b$, and $\tau$, where (see Equations~(\ref{eq40}))
\begin{equation}
a \equiv \frac{A_0}{D_0} \ , \qquad
b \equiv \frac{B_0}{D_0} \ , \qquad
\tau \equiv \frac{R^2 q B D_0}{m_e c^3} \ ,
\label{eq61}
\end{equation}
and use these definitions to rewrite the steady-state transport equation as
\begin{align}
\frac{1}{D_0}\frac{\partial \sgreen}{\partial t} =& 0 = \frac{1}{x^2} \frac{d}
{d x} \left(x^4\frac{d \sgreen}{d x}
- a x^3 \sgreen + b x^4 \sgreen \right) - \frac{x \sgreen}{\tau} \nonumber \\
&+ \frac{\dot N_0 \delta(x-x_0)}{4 \pi D_0 (m_e c)^3x_0^2} \ .
\label{eq62}
\end{align}
\subsection{Steady-state Electron Green's Function}

Noting the similarity between Equations~(\ref{eq44}) and (\ref{eq62}), and recognizing that $\sgreen$ must be continuous at $x=x_0$, we can write the global solution for $\sgreen$ as (cf. Equation~(\ref{eq47}))
\begin{equation}
\sgreen(x) = H_0 \, e^{-b x/2 } x^{-2 + a/2}
M_{\lambda,\sigma}(b \xmin) \,
W_{\lambda,\sigma} (b \xmax) \ ,
\label{eq63}
\end{equation}
where $\xmin$ and $\xmax$ are defined by Equations~(\ref{eq48}), and the parameters $\lambda$ and $\sigma$ are given by
\begin{equation}
\lambda = 2-\frac{1}{b\tau} + \frac{a}{2} \ , \qquad
\sigma = \frac{a+3}{2} \ .
\label{eq64}
\end{equation}

The steady-state Green's function $\sgreen$ is continuous at the injection momentum $x_0$, and it displays a derivative jump with a magnitude that is obtained by integrating Equation~(\ref{eq62}) in a small region around the injection momentum. The result obtained is
\begin{equation}
\lim_{\delta \to 0} \left[\frac{d\sgreen}{dx} \right]_{x_0+\delta}
- \lim_{\delta \to 0} \left[\frac{d\sgreen}{dx} \right]_{x_0-\delta}
= - \frac{\dot{N}_0}{4 \pi D_0 (m_e c)^3 x_0^4} \ .
\label{eq65}
\end{equation}
Using Equation~(\ref{eq63}) to substitute for $\sgreen$ in Equation~(\ref{eq65}), we find that
\begin{align}
H_0 &\, e^{-b x_0/2 } x_0^{-2 + a/2} \, b \,
\big[M_{\lambda,\sigma}(bx_0) W'_{\lambda,\sigma}(bx_0)
- W_{\lambda,\sigma}(bx_0) \nonumber \\
&\times M'_{\lambda,\sigma}(bx_0)\big]
= - \frac{\dot N_0}{4 \pi D_0 (m_e c)^3 x_0^4} \ .
\label{eq66}
\end{align}
Utilizing Equation~(\ref{eq51}) for the Wronskian and solving for $H_0$ yields
\begin{equation}
H_0 = \frac{\dot N_0 e^{b x_0/2}}
{4 \pi b D_0 (m_e c)^3 x_0^{2+a/2}}
\frac{\Gamma(\sigma-\lambda+1/2)}{\Gamma(1+2\sigma)} \ .
\label{eq67}
\end{equation}
Combining Equations~(\ref{eq63}) and (\ref{eq67}), we find that the closed-form solution for the steady-state electron Green's function, $\sgreen(x)$, is given by
\begin{align}
\sgreen(x) =& \frac{\dot N_0 e^{b x_0/2}}
{4 \pi b D_0 (m_e c)^3 x_0^{2+a/2}} \frac{\Gamma(\sigma-\lambda+1/2)}
{\Gamma(1+2\sigma)} \, e^{-b x/2 } \nonumber \\
&\times x^{-2 + a/2}
M_{\lambda,\sigma}(b x_{\rm min}) \,
W_{\lambda,\sigma} (b x_{\rm max}) \ ,
\label{eq68}
\end{align}
where the constants $\lambda$ and $\sigma$ are evaluated using Equations~(\ref{eq64}) and $x_{\rm min}$ and $x_{\rm max}$ are defined by Equations~(\ref{eq48}). Equation~(\ref{eq68}) gives the steady-state electron distribution function in the co-moving blob frame resulting from the continual injection of monoenergetic electrons, which is used to derive the electron number distribution (see Equation~(\ref{eq76})). In Section~6, we will use Equation~(\ref{eq68}) to compute the synchrotron spectrum emitted by the steady-state distribution of relativistic electrons during the peak of the X-ray flare, which is plotted in Figure~2.

\subsection{Fokker-Planck Equation}

It is instructive to recast the steady-state transport equation in Fokker-Planck form because the resulting Fokker-Planck coefficients provide additional insight into the physical process involved in the particle transport scenario under consideration here. To begin, we define the steady-state electron number distribution, $\Ngreen(x)$, using (see Equation~(\ref{eq54}))
\begin{equation}
\Ngreen(x) = 4 \pi \, (m_ec)^3x^2\sgreen(x) \ ,
\label{eq69}
\end{equation}
which is related to the (constant) total number of steady-state electrons in the blob, ${\cal N}_e^{\rm S}$, via (see Equation~(\ref{eq55}))
\begin{equation}
{\cal N}_e^{\rm S} = \int_{0}^{\infty} \Ngreen(x) \, dx \ .
\label{eq70}
\end{equation}

Using Equation~(\ref{eq69}) to substitute for $\sgreen$ is Equation~(\ref{eq62}) and rearranging the resulting expression, we obtain
\begin{align}
\frac{1}{D_0} \frac{\partial \Ngreen}{\partial t} =& 0 = \frac{d^2}{d x^2} \left(\Ngreen x^2 \right)
- \frac{d}{d x} \left[\left(4x+ax-bx^2 \right)\Ngreen \right] \nonumber \\
&- \frac{\Ngreen x}{\tau}
+ \frac{\dot{N}_0 \delta(x-x_0)}{D_0} = 0 \ .
\label{eq71}
\end{align}
This equation can be expressed in Fokker-Planck form by writing
\begin{align}
\frac{\partial \Ngreen}{\partial t} =& \frac{\partial^2}{\partial x^2} \left( \frac{1}{2}
\frac{d \sigma^2}{d t} \Ngreen \right) - \frac{\partial}{\partial x}
\left( \left< \frac{d x}{d t} \right> \Ngreen \right) \nonumber \\
&- \frac{\Ngreen x D_0}{\tau}
+ \dot{N}_0 \delta(x-x_0) = 0 \ ,
\label{eq72}
\end{align}
where the ``broadening coefficient'' is given by
\begin{equation}
\frac{1}{2} \frac{d \sigma^2}{d t} = D_0 \, x^2 \ ,
\label{eq73}
\end{equation}
and the ``drift coefficient,'' describing the mean net acceleration rate, is given by
\begin{equation}
\left< \frac{d x}{d t} \right> = D_0 (4x+ax-bx^2) \ .
\label{eq74}
\end{equation}
The drift coefficient represents the mean electron acceleration rate, which vanishes when acceleration is balanced by synchrotron losses. Hence, we can estimate the equilibrium Lorentz factor for the electrons, $x_{\rm eq}$, by setting $<\!d x/d t\!>=0$. The result obtained is
\begin{equation}
x_{\rm eq}=\frac{a+4}{b} \ .
\label{eq75}
\end{equation}
This expression will be used in Section~6, when we apply our model to a specific astrophysical source.

In calculating the X-ray spectrum of the flare produced via synchrotron emission, it will be convenient to work in terms of the steady-state electron number distribution, $\Ngreen$, measured by an observer in the co-moving frame of the outflowing plasma blob. We can obtain an expression for $\Ngreen$ by combining Equations~(\ref{eq68}) and (\ref{eq69}), which yields
\begin{align}
\Ngreen(x) =& \frac{\dot N_0 e^{b x_0/2}}
{b D_0 x_0^{2+a/2}} \frac{\Gamma(\sigma-\lambda+1/2)}
{\Gamma(1+2\sigma)} \, e^{-b x/2 } x^{a/2} \nonumber \\
&M_{\lambda,\sigma}(b x_{\rm min}) \,
W_{\lambda,\sigma} (b x_{\rm max}) \ ,
\label{eq76}
\end{align}
where $x_{\rm max}$ and $x_{\rm min}$ are defined by Equations~(\ref{eq48}) and the constants $\lambda$ and $\sigma$ are evaluated using Equations~(\ref{eq64}). Equation~(\ref{eq76}) is interpreted as the co-moving electron distribution occurring during the peak of the X-ray fare, when a balance is achieved between particle acceleration, losses, injection, and escape. The electron distribution is plotted in Figure~2, and further discussed in Section~7.1.

\section{SYNCHROTRON EMISSION}

The synchrotron spectrum radiated by electrons with Lorentz factor $\gamma$ displays a peak at the photon energy (Rybicki \& Lightman 1979)
\begin{equation}
\epsilon(x) = \xi \frac{B}{B_c} \, \gamma^2 m_e c^2 \ ,
\label{eq77}
\end{equation}
where $\xi$ is an order-unity constant, and $B_c=(2\pi m_e^2 c^3)/(e h) \approx 4.41 \times 10^{13}\,$G denotes the critical field strength. We will set $\xi=1$ in our applications. The Lorentz factor $\gamma$ is related to the dimensionless momentum $x$ via $x=\sqrt{\gamma^2-1}$, so we can set $x=\gamma$ for the ultrarelativistic electrons of interest here. An exact calculation of the synchrotron spectrum requires numerical integration, but we can obtain reasonably accurate results using the $\delta$-function approximation, in which the number of photons generated per unit time per unit energy due to synchrotron emissivity in the frame of the blob is given by (e.g., Dermer \& Menon 2009)
\begin{equation}
\dot N^{\rm syn}_\epsilon(\epsilon,t) = \frac{2}{3} \, c \,
\sigmaT U_B x^3 \epsilon^{-2} N_e(x,t)
\ \ \propto \ \ {\rm s^{-1}\,erg^{-1}}
\ ,
\label{eq78}
\end{equation}
where $N_e(x,t)$ is the number distribution of the electrons, and the dimensionless momentum $x$ is given in terms of $\epsilon$ by
\begin{equation}
x(\epsilon) = \left(\frac{B_c}{B} \frac{\epsilon}{\xi m_e c^2}\right)^{1/2}
\ ,
\label{eq79}
\end{equation}
which is obtained by setting $\gamma=x$ in Equation~(\ref{eq77}).

In order to connect our theory with the observational data, we will need to relate the synchrotron emissivity given by Equation~(\ref{eq78}) to the observed X-ray spectrum. It is convenient to introduce the specific luminosity function, $\mathscr{L}(\epsilon,t)$, defined by
\begin{equation}
\mathscr{L}(\epsilon,t) \equiv \epsilon L_\epsilon(\epsilon,t) \ \propto \ {\rm erg \ s^{-1}} \ ,
\label{eq80}
\end{equation}
where $L_\epsilon$ is the specific luminosity, which is related to the synchrotron emissivity, $\dot N^{\rm syn}_\epsilon$, via (Dermer \& Menon 2009)
\begin{equation}
L_\epsilon(\epsilon,t) = \epsilon \dot N^{\rm syn}_\epsilon(\epsilon,t)
\ \propto \ {\rm s^{-1}} \ .
\label{eq81}
\end{equation}
We can combine Equations~(\ref{eq78}), (\ref{eq80}), and (\ref{eq81}) to show that in the $\delta$-function approximation,
\begin{equation}
\mathscr{L}(\epsilon,t) = \frac{2}{3} \, c \, \sigmaT
U_B \, x^3 N_e(x,t) \ \propto \ {\rm erg \ s^{-1}} \ .
\label{eq82}
\end{equation}
This relation allows us to compute the specific luminosity function, $\mathscr{L}$, based on knowledge of the electron number distribution, $N_e$. Since the emitting electrons are located in the outflowing plasma blob, we must interpret $\mathscr{L}$ and $N_e$ in Equation~(\ref{eq82}) as co-moving distributions. This is further discussed below.

\subsection{Transformation Between Frames}

In all of the preceding analysis, we have been working in the co-moving frame of the plasma blob, which travels outward through the jet with velocity $v=\beta c$ and bulk Lorentz factor $\Gamma = (1-\beta^2)^{-1/2}$. The blob has a Doppler factor, $\delta_{\rm D} = [\Gamma(1-\beta \cos \theta)]^{-1}$, where $\theta$ is the angle between the jet axis and the line of sight to the observer. In order to make a connection between the radiation emitted in the co-moving frame and the spectrum measured by a distant observer, we must apply a transformation to account for Lorentz invariance combined with cosmological effects. We can transform between the photon energies measured in the two frames using
\begin{equation}
\frac{\epsilon}{\epsilon'} = \frac{\delta_{\rm D}}{1+z} \ ,
\label{eq83}
\end{equation}
where primes denote quantities measured in the co-moving frame and $z$ is the cosmological redshift. Likewise, time dilation implies that the Fourier frequencies and the time intervals in the two frames are related
via (Finke \& Becker 2014)
\begin{equation}
\frac{\omega'}{\omega} = \frac{t}{t'} = \frac{1+z}{\delta_{\rm D}} \ .
\label{eq84}
\end{equation}
It follows that for a given observed photon energy, $\epsilon$, the co-moving Lorentz factor, $x'$, is given by (cf.
Equation~(\ref{eq77}))
\begin{equation}
x' = \left(\frac{\epsilon}{m_e c^2} \frac{B_c}{B\xi}
\frac{1+z}{\delta_{\rm D}}\right)^{1/2}
\ .
\label{eq85}
\end{equation}
Another useful quantity is the equilibrium Lorentz factor, corresponding to a balance between acceleration and losses, which can be written in the co-moving frame notation as (see Equation~(\ref{eq75}))
\begin{equation}
x'_{\rm eq}=\frac{a+4}{b} \ .
\label{eq86}
\end{equation}

Equation~(\ref{eq82}) gives the synchrotron spectrum generated in the co-moving frame of the outflowing plasma blob, which can be written in the primed notation as
\begin{equation}
\mathscr{L}'(\epsilon',t') = \frac{2}{3} \, c \, \sigmaT
U_B \, x'^3 N'_e(x',t') \ \ \propto \ \ {\rm erg \ s^{-1}} \ .
\label{eq87}
\end{equation}
The specific luminosity function, $\mathscr{L}$, transforms between the observer frame (unprimed) and the co-moving frame (primed) according to (Dermer \& Menon 2009)
\begin{equation}
\mathscr{L}(\epsilon,t) = \delta_{\rm D}^4 \, \mathscr{L}(\epsilon',t')
\ ,
\label{eq88}
\end{equation}
where the energies, $\epsilon$ and $\epsilon'$, and times, $t$ and $t'$, are related via Equations~(\ref{eq83}) and (\ref{eq84}), respectively. The related specific flux function, $\mathscr{F}$, is defined by
\begin{equation}
\mathscr{F}(\epsilon,t) \equiv \frac{1}{4 \pi d_L^2} \, \mathscr{L}(\epsilon,t) \ ,
\label{eq89}
\end{equation}
where $d_L$ denotes the luminosity distance. Equation~(\ref{eq89}) can also be written in the equivalent form
\begin{equation}
\mathscr{F}(\epsilon,t) = \epsilon F_\epsilon(\epsilon,t)
= \frac{1}{4 \pi d_L^2} \, \epsilon L_\epsilon(\epsilon,t)
\ ,
\label{eq90}
\end{equation}
where $F_\epsilon$ is the observed specific flux and $L_\epsilon$ is the observed specific luminosity introduced in Equation~(\ref{eq80}). Combining Equations~(\ref{eq87}), (\ref{eq88}), (\ref{eq89}), and (\ref{eq90}), we obtain
\begin{equation}
\mathscr{F}(\epsilon,t) = \frac{\delta_{\rm D}^4}{6 \pi d_L^2}
\, c \, \sigmaT U_B \, x'^3 N'_e(x',t')
\ ,
\label{eq91}
\end{equation}
where $x'$ is given by Equation~(\ref{eq85}) and $N'_e$ represents the co-moving electron number distribution.

\subsection{Steady-state X-ray Spectrum}

During the peak of the X-ray flare, the electrons will possess an approximate equilibrium distribution if there is enough time to establish a balance between the various competing processes, as discussed in Section~7.1. In the calculations of the peak flare X-ray spectrum presented in Section~6, we will assume that the electrons in the blob have the steady-state distribution given by Equation~(\ref{eq76}), which is written in the co-moving frame notation as
\begin{align}
N_e'(x')=& \frac{\dot N_0' e^{b(x'_0-x')/2}}
{b D_0 {x'}_0^2} \frac{\Gamma(\sigma-\lambda+1/2)}
{\Gamma(1+2\sigma)} \, \left(\frac{x'}{x'_0}\right)^{a/2} \nonumber \\
& M_{\lambda,\sigma}(b x'_{\rm min}) \,
W_{\lambda,\sigma} (b x'_{\rm max}) \ ,
\label{eq92}
\end{align}
where
\begin{equation}
x'_{\rm min} \equiv \min(x',x'_0) \ , \qquad
x'_{\rm max} \equiv \max(x',x'_0) \ ,
\label{eq93}
\end{equation}
and
\begin{equation}
\lambda = 2-\frac{1}{b\tau} + \frac{a}{2} \ , \qquad
\sigma = \frac{a+3}{2} \ .
\label{eq94}
\end{equation}
Here, $x'_0$ denotes the Lorentz factor of the injected electrons as measured in the co-moving frame. Using Equation~(\ref{eq92}) to substitute for $N_e'$ in Equation~(\ref{eq91}) yields the final form for the observed steady-state specific flux function, denoted by $\mathscr{F}^{\rm S}(\epsilon)$. We obtain
\begin{align}
\mathscr{F}^{\rm S}(\epsilon) =& \frac{\delta_{\rm D}^4}{6 \pi d_L^2}
\frac{\dot N_0' \, c \, \sigmaT U_B x'^3 e^{b(x'_0-x')/2}}
{b D_0 {x'}_0^2} \frac{\Gamma(\sigma-\lambda+1/2)}
{\Gamma(1+2\sigma)} \nonumber \\
&\times \left(\frac{x'}{x'_0}\right)^{a/2}
M_{\lambda,\sigma}(b x'_{\rm min}) \,
W_{\lambda,\sigma} (b x'_{\rm max})
\ ,
\label{eq95}
\end{align}
where $x'$ is given in terms of the observed photon energy $\epsilon$ using Equation~(\ref{eq85}), and $\lambda$ and $\sigma$ are computed using Equations~(\ref{eq94}). We will use Equation~(\ref{eq95}) to calculate the observed specific flux $\mathscr{F}^{\rm S}(\epsilon)\equiv \epsilon F_\epsilon$ when comparing our model predictions with the X-ray spectrum observed during the peak of the flare, which is plotted in Figure~2.

\subsection{Fourier Transformation and Time Lags}

To apply our theoretical model to the computation of X-ray time lags, we must develop an expression for the Fourier transform of the time-dependent specific flux function, $\mathscr{F}(\epsilon,t) \equiv \epsilon F_\epsilon$ (see Equation~(\ref{eq89})). We define the Fourier transform of $\mathscr{F}$ using
\begin{equation}
\tilde{\mathscr{F}}(\epsilon,\omega) \equiv \int_{-\infty}^\infty
e^{i \omega t} \, \mathscr{F}(\epsilon,t) \, dt \ .
\label{eq96}
\end{equation}
The general relation between $\mathscr{F}$ and the co-moving electron distribution $N_e'$ is given by Equation~(\ref{eq91}). Using Equation~(\ref{eq91}) to substitute for $\mathscr{F}$ in
Equation~(\ref{eq96}) yields
\begin{equation}
\tilde{\mathscr{F}}(\epsilon,\omega) =
\frac{\delta_{\rm D}^4}{6 \pi d_L^2}
\, c \, \sigmaT U_B x'^3 \int_{-\infty}^\infty
e^{i \omega t} \, N'_e(x',t') \, dt
\ ,
\label{eq97}
\end{equation}
where $x'$ is computed using Equation~(\ref{eq85}).

Based on the reciprocal relation between time and Fourier frequency evidenced by the frame transformations (Equation~(\ref{eq84})), we can make the change of variables from $t$ to $t'$ by writing
\begin{equation}
\omega \, t = \omega' \, t' \ ,
\label{eq98}
\end{equation}
so that Equation~(\ref{eq97}) now becomes
\begin{equation}
\tilde{\mathscr{F}}(\epsilon,\omega) =
\frac{1+z}{\delta_{\rm D}} \frac{\delta_{\rm D}^4}{6 \pi d_L^2}
\, c \, \sigmaT U_B x'^3
\int_{-\infty}^\infty e^{i \omega' t'} \, N'_e(x',t') \, dt'
\ ,
\label{eq99}
\end{equation}
or, equivalently,
\begin{equation}
\tilde{\mathscr{F}}(\epsilon,\omega) =
\frac{(1+z) \delta_{\rm D}^3}{6 \pi d_L^2}
\, c \, \sigmaT U_B x'^3
\tilde N'_e(x',\omega') \ ,
\label{eq100}
\end{equation}
where the co-moving electron Fourier transform is defined by
\begin{equation}
\tilde N'_e(x',\omega') \equiv \int_{-\infty}^\infty
e^{i \omega' t'} \, N'_e(x',t') \, dt'
\ .
\label{eq101}
\end{equation}

In order to proceed, we need to evaluate the co-moving electron Fourier transform, $\tilde N_e'$, which is written in the co-moving frame notation as (see Equation~(\ref{eq57}))
\begin{align}
\tilde N'_e(x',\omega') &= \frac{N_0 e^{i \omega' t'_0} e^{b(x'_0-x')/2}}
{b D_0 {x'}_0^2} \frac{\Gamma(\mu-\kappa+1/2)}{\Gamma(1+2\mu)}
\nonumber \\ 
&\times \left(\frac{x'}{x'_0}\right)^{a/2}
M_{\kappa,\mu}(b x'_{\rm min}) \,
W_{\kappa,\mu} (b x'_{\rm max})
\ .
\label{eq102}
\end{align}
The final result for the Fourier transform of the observed specific flux, $\tilde{\mathscr{F}}$, is obtained by combining Equations~(\ref{eq100}) and (\ref{eq102}), which yields
\begin{align}
&\tilde{\mathscr{F}}(\epsilon,\omega) =
\frac{(1+z) \delta_{\rm D}^3}{6 \pi d_L^2}
\frac{N_0 \, c \, \sigmaT U_B x'^3 e^{i\omega't_0'} e^{b(x'_0-x')/2}}
{b D_0 {x'}_0^2} \hfil \nonumber \\
&\times \frac{\Gamma(\mu-\kappa+1/2)}{\Gamma(1+2\mu)}
\, \left(\frac{x'}{x'_0}\right)^{a/2}
M_{\kappa,\mu}(b x'_{\rm min}) \,
W_{\kappa,\mu} (b x'_{\rm max}) \ ,
\label{eq103}
\end{align}
where $x'$ is calculated from the observed photon energy, $\epsilon$, using Equation~(\ref{eq85}),
$x'_{\rm min}$ and $x'_{\rm max}$ are defined by Equations~(\ref{eq93}), $t_0'$ is the injection time in the co-moving frame, and $\kappa$ and $\mu$ are evaluated using
\begin{equation}
\kappa = 2-\frac{1}{b\tau} + \frac{a}{2} \ , \qquad
\mu = \sqrt{\frac{(a+3)^2}{4} - \frac{i \omega}{D_0}
\frac{1+z}{\delta_{\rm D}}} \ .
\label{eq104}
\end{equation}
Here, $\omega$ denotes the observer-frame Fourier frequency, which is related to the co-moving frequency, $\omega'$, via Equation~(\ref{eq84}).

Equations~(\ref{eq103}) and (\ref{eq104}) give the closed-form solution for the Fourier transform, $\tilde{\mathscr{F}}$, of the observed time-dependent specific specific flux, $\mathscr{F}(\epsilon,t)\equiv\epsilon F_\epsilon$, stated in terms of the observed X-ray energy, $\epsilon$, and the observed Fourier frequency, $\omega$. These expressions can therefore be used to generate theoretical values for the soft and hard Fourier transforms, $G_s(\omega)$ and $G_h(\omega)$, corresponding to the soft and hard X-ray channel energies, $\epsilon_s$ and $\epsilon_h$, respectively, by writing
\begin{equation}
G_s(\omega) = \tilde{\mathscr{F}}(\epsilon_s,\omega) \ , \qquad
G_h(\omega) = \tilde{\mathscr{F}}(\epsilon_h,\omega)
\ .
\label{eq105}
\end{equation}
The Fourier transforms $G_s(\omega)$ and $G_h(\omega)$ are then utilized to compute the X-ray time lags using Equations~(\ref{eq1}) and (\ref{eq4}). This procedure allows us to compute the theoretical time lag between any two selected X-ray channel energies, $\epsilon_s$ and $\epsilon_h$, for any value of the Fourier frequency $\omega$. In Section~6 we will make an application of this method to Mrk~421. By comparing the predicted time lags with the observational data, we can test the theory, and also constrain the model parameters.

\section{APPLICATION TO MRK 421}

The goal of this paper is to develop a theoretical framework based on first-principles physical concepts that can be used to model the transport and acceleration of relativistic electrons in blazar jets. In particular, we are interested in determining whether a single transport model can simultaneously account for both the X-ray time lags, and the shape of the peak flare X-ray spectrum. In this section, we will use the 1998 April 21 X-ray flare from Mrk~421 as a sample application. The analysis proceeds along two separate tracks, with the first focusing on the interpretation of the X-ray time lags, and the second focusing on the interpretation of the peak flare X-ray spectrum. The two tracks yield two respective sets of model parameters that can be compared and synthesized to deduce the nature of the physics occurring in the plasma blob during the X-ray flare.

\subsection{X-ray Time Lags}

Zhang (2002) analyzed {\it Beppo}SAX data collected during the 1998 April 21 flare of Mrk~421. He obtained hard time lags of about one hour by Fourier transforming the data in two energy windows and then applying our Equations~(\ref{eq1}) and (\ref{eq4}). Similar results were obtained by Fossati et al. (2000a), although these authors employed a different method, based on the discrete correlation function, rather than utilizing Fourier transformation. We will therefore compare our model predictions with the Fourier time lags obtained by Zhang (2002). Computation of the time lags using our model requires the specification of input values for the co-moving blob radius $R'$, the Doppler factor $\delta_{\rm D}$, the magnetic field $B$, the redshift $z$, and the luminosity distance $d_L$. The observed redshift $z=0.031$ for Mrk~421 gives a luminosity distance $d_L = 4.2 \times 10^{26}\,$cm, assuming $H_0=70\,\rm km\,s^{-1}\,Mpc^{-1}$, $\Omega_m=0.3$, and $\Omega_{\lambda}=0.7$. The remaining parameters $R'$, $\delta_{\rm D}$, and $B$ can be estimated using the detailed spectral analyses carried out by Fossati et al. (2000b), Abdo et al. (2011), and Finke et al. (2008).

It is important to note that the blob radius $R'$ is constrained by the time lag, $\delta t'$, measured in the co-moving frame, which must exceed the light-crossing time of the blob in order to avoid causality violations. We can therefore write
\begin{equation}
\delta t' > \frac{R'}{c} \ .
\label{eq106}
\end{equation}
Due to the Doppler boost, combined with the cosmological redshift, the time lag measured in the observer frame, $\delta t$, is related to $\delta t'$ via (see Equation~(\ref{eq84}))
\begin{equation}
\delta t = \delta t' \left(\frac{1+z}{\delta_{\rm D}}\right) \ .
\label{eq107}
\end{equation}
Combining Equations~(\ref{eq106}) and (\ref{eq107}) yields a causality constraint on $R'$ and $\delta_{\rm D}$, given by (Diltz \& B\"ottcher 2014; Abdo et al. 2011)
\begin{equation}
\delta t \, > \left(\frac{1+z}{\delta_{\rm D}}\right) \frac{R'}{c} \ .
\label{eq108}
\end{equation}
Out of the total of 10 models for Mrk~421 considered by Fossati et al. (2000b), Abdo et al. (2011), and Finke et al. (2008), only four satisfy the causality constraint given by Equation~(\ref{eq108}), assuming $\delta t \sim 1\,$hour. Here, we will focus on the ``green'' leptonic model listed in Table~4 from Abdo et al. (2011), with parameter values $B=0.082\,$G, $R'=5.3\times10^{15}\,$cm, and $\delta_{\rm D}=50$, and therefore these are the values adopted in our analysis.

Once the values of $B$, $R'$, $\delta_{\rm D}$, $z$, and $d_L$ are specified for Mrk~421, the remaining free parameters in our model are the dimensionless shock acceleration/adiabatic loss parameter $a$, the dimensionless synchrotron loss parameter $b$, and the injected Lorentz factor $x'_0$, which is measured in the co-moving frame. We vary the values of $x'_0$, $a$, and $b$ so as to achieve good qualitative agreement with the time lag data reported by Zhang (2002) as a function of the Fourier frequency $\nu_f$. The value of $D_0$ is internally computed by combining Equations (\ref{eq34}) and (\ref{eq40}) to obtain
\begin{equation}
D_0 = \frac{\sigmaT B^2}{6 \pi m_e c \, b} \ .
\label{eq109}
\end{equation}
Additionally, we compute the value of $\tau$, the dimensionless escape constant, according to the definition in Equation~(\ref{eq40})
\begin{equation}
\tau = \frac{R'^2 q B D_0}{m_e c^3} \ .
\label{eq110}
\end{equation}

The energy windows used by Zhang (2002) extended from 0.1-2.0\,keV for the soft energy channel, and from 2.0-10.0\,keV for the hard energy channel. However, our model requires the specification of precise values for the hard and soft channel energies, $\epsilon_h$ and $\epsilon_s$, in order to generate theoretical predictions for the time lags, and therefore we need to extract two characteristic energies from Zhang's hard and soft windows. One can imagine a variety of different averaging schemes, but the most obvious possibility is to select the channel-center energies from Zhang's two windows, which yields $\epsilon_s=1.05\,$keV and $\epsilon_h=6.00\,$keV. We will utilize these energies as our primary values for $\epsilon_h$ and $\epsilon_s$, but we will also examine two alternative calculations based on different values for $\epsilon_h$ and $\epsilon_s$ in Section~8.3.

In Figure~1, we plot the X-ray time lags computed using Equations~(\ref{eq1}) and (\ref{eq4}), with the hard and soft Fourier components $G_h(\omega)$ and $G_s(\omega)$ evaluated using Equations~(\ref{eq105}). We set the hard and soft channel energies $\epsilon_h$ and $\epsilon_s$ equal to Zhang's channel-center energies, 6\,keV and 1.05\,keV, respectively. In the sign convention we adopt, a positive time lag is obtained when the hard X-ray signal lags the soft signal. The results plotted in Figure~1 therefore indicate that we obtain a hard time lag at all Fourier frequencies below the very sharp turnover at frequency $\nu_f \sim 10^{-4.05}\,$Hz, where the lag turns negative (soft). Figure~1 also includes the time lags derived by Zhang (2002) based on analysis of the {\it Beppo}SAX data obtained during the 1998 April 21 flare of Mrk~421. It is apparent from Figure~1 that our theoretical model, based on the impulsive injection of monoenergetic electrons, is able to qualitatively reproduce the time lags observed from Mrk~421 during the 1998 April 21 flare as a function of the Fourier frequency, $\nu_f = 2 \pi \omega$, including the production of hard time lags, and the appearance of a very sharp transition to a soft lag above the frequency $\nu_f \sim 10^{-4.05}\,$Hz. As far as we are aware, this is the first time that the time lag observations reported by Zhang (2002) have been explained using any physics-based model. The physical significance of the time-lag results plotted in Figure~1 is further discussed in Section~8.2.

The dimensionless theoretical parameters used to generate the time lags plotted in Figure~1 (corresponding to the ``primary'' channel energy values $\epsilon_s=1.05\,$keV and $\epsilon_h=6.00\,$keV) are $a = 40.0$, $b = 7.94 \times 10^{-5}$, and $x'_0 = 2.55 \times 10^5$, and the corresponding values for the parameters $D_0$ and $\tau$ obtained using Equations~(\ref{eq109}) and (\ref{eq110}) are $D_0 = 1.09 \times 10^{-7}\,{\rm s}^{-1}$ and $\tau = 4.93 \times 10^{9}$, respectively (see Table~1). The large value of $x'_0$ indicates that the impulsively injected seed electrons are already highly relativistic, and therefore they cannot be picked up from the thermal electron distribution in the plasma. We therefore hypothesize that the high-energy seed electrons are generated via magnetic reconnection in the vicinity of the shock waves (or waves) inside the plasma blob (Nalewajko et al. 2011; Giannios 2013). The value the equilibrium Lorentz factor, $x'_{\rm eq}$, computed using Equation~(\ref{eq86}), is $x'_{\rm eq} = 5.54 \times 10^5$, which exceeds the injected Lorentz factor $x'_0$. This indicates that the injected electrons experience further acceleration due to interactions with the shock(s) and the MHD waves, which is consistent with the large positive value of $a$ we obtain, suggesting that shock acceleration overwhelms adiabatic losses during the rapid transients that produce the observed X-ray time lags. We provide additional discussion of the physical interpretation of the model parameters in Section~7.

\subsection{Peak Flare X-ray Spectrum}

Our goal is to produce an integrated model that can simultaneously account for the time lags and the X-ray spectrum observed during the flare. Hence in this section we will compute the steady-state X-ray spectrum generated by electrons that are continuously injected into the jet to see how closely it resembles the spectrum observed during the peak of the X-ray flare. In particular, we are interested in determining whether the flare X-ray spectrum can be reproduced using a set of theory parameters that are similar to the parameters used to model the observed X-ray time lags, as discussed in Section~6.1.

The X-ray spectrum observed during the 1998 April 21 flare was reported and discussed by Fossati et al. (2000b). Those observations are contemporaneous with the time-lag data analyzed by Zhang (2002), so it is especially interesting to apply our model to the interpretation of the Fossati et al. (2000b) X-ray spectral
data. In our computation of the X-ray spectrum, we again adopt the values $B=0.082\,$G, $R'=5.3\times10^{15}\,$cm, and $\delta_{\rm D}=50$ taken ``green'' leptonic model considered by Abdo et al. (2011). With these parameters set, along with the redshift $z = 0.031$ and the luminosity distance $d_L = 4.2 \times 10^{26}\,$cm, we vary the remaining theory parameters $a$, $b$, $x'_0$, and $\dot N'_0$ so as to achieve good qualitative agreement with the X-ray spectrum observed during the peak of the flare, as reported in Figure~3 from Fossati et al. (2000b).

In Figure~2, we plot the steady-state specific flux, $\mathscr{F}^{\rm S} \equiv \nu F_\nu$, evaluated as a function of the photon frequency $\nu$ using Equation~(\ref{eq95}). Figure~2 also includes the peak X-ray spectrum observed during the 1998 April 21 flare, taken from Figure~3 in Fossati et al. (2000b). The dimensionless model parameter values used to generate the X-ray spectrum in Figure~2 are $a = -3.30$, $b = 1.02 \times 10^{-5}$, $x'_0=2$, and $\dot N'_0 = 2.82 \times 10^{34}\,\rm s^{-1}$, and the associated values of $D_0$ and $\tau$ computed using Equations~(\ref{eq109}) and (\ref{eq110}) are $D_0 = 8.49 \times 10^{-7}\,{\rm s}^{-1}$ and $\tau = 3.83 \times 10^{10}$, respectively (see Table~1). The equilibrium Lorentz factor computed using Equation~(\ref{eq86}) is $x'_{\rm eq} = 6.84 \times 10^4$, and the resulting X-ray spectrum is quite insensitive to the value for the injected Lorentz factor, $x'_0$, because the memory of the injected electron energy is lost as a result of multiple interactions with the MHD waves. We note that the theoretical X-ray spectrum agrees fairly well with the observational data.

The parameters used to compute the steady-state spectrum are similar to those used to generate the time lags, but there are some important differences. For example, the Lorentz factor of the continually injected electrons is $x'_0=2$, which is far smaller than the value $x'_0 \sim 10^5$ obtained in the time lag calculation. This implies that the steady-state electron distribution may result from the injection of mildly relativistic thermal electrons picked up from the thermal distribution in the blob. On the other hand, the very high energy of the injected electrons in the time lag calculation implies that a very energetic process is producing those particles, such as impulsive magnetic reconnection, probably occurring in the vicinity of a shock wave in the blob (Nalewajko et al. 2011; Sironi et al. 2015).

Another important difference is that the value of $a$ is negative in the spectrum calculation, whereas it is positive in the time lag calculation discussed in Section~6.1. The negative value of $a$ obtained in the spectrum calculation indicates that losses due to adiabatic expansion dominate over gains due to shock acceleration during the formation of the peak flare X-ray spectrum. The large value of $x'_{\rm eq}$ therefore implies that the dominant form of particle acceleration is second-order Fermi (stochastic) acceleration due to interactions with MHD waves in the plasma blob, which is able to accelerate the seed electrons to highly relativistic energies.

In Figure~3 we plot the steady-state distribution of the electrons responsible for producing the model X-ray spectrum plotted in Figure~2. The electron number distribution is plotted as a function of the Lorentz factor $\gamma'$ (or equivalently $x'$), as measured by an observer in the co-moving jet frame, computed using Equation~(\ref{eq92}). The model parameters $a$, $b$, $B$, $x'_0$, $\dot N'_0$, $\delta_{\rm D}$, $z$, and $d_L$ are identical to those used to calculate the X-ray spectrum plotted in Figure~2. We note that the electron distribution obtained is similar in magnitude and shape to those computed using the power-law method employed by Finke et al. (2008). However, we emphasize that our results for the electron distribution are obtained using a first-principles physical model, in contrast to an ad hoc power-law fit. The electron number distribution extends up to a Lorentz factor of $\sim 10^5$, as expected, since the equilibrium Lorentz factor in this case is $x'_{\rm eq} = 6.84 \times 10^4$. 

\section{PARAMETER CONSTRAINTS}

The synthesis of the spectral and timing information facilitated by the new model provides a powerful new tool for probing the detailed physics occurring in the blazar jet. In this section, we analyze the validity of the key assumptions underlying our model, and we also connect the theoretical parameters more directly with the physical properties of the jet.

\subsection{Equilibration Timescale}

In our calculation of the X-ray spectrum observed during the peak of the 1998 April 21 flare from Mrk~421, we have assumed that electrons comprise a steady-state distribution. This implies that the particles have achieved, at least approximately, an equilibrium between the competing processes of first- and second-order Fermi acceleration, synchrotron and adiabatic losses, and particle injection and escape. It is important to examine the validity of this assumption.

We note  that transport coefficients themselves are not likely to vary on the same timescales as the observed emission, which displays a hard time lag of roughly one hour. This conclusion is based on the fact that the short-timescale variability involves only $\sim 10\%$ of the total X-ray flux amplitude, and therefore the dynamical structure of the shock, and the field of MHD waves, is not likely to be strongly perturbed, when averaged over the volume of the blob.

If the transport coefficients are not time-dependent, then the time required for equilibrium to be established should be comparable to the synchrotron loss timescale, which is the dominant energy loss timescale for the problem. The characteristic synchrotron loss timescale in the co-moving frame of the outflowing plasma blob can be estimated by setting $E = x'_{\rm eq} m_e c^2$ in Equation~(\ref{eq11}), obtaining
\begin{equation}
t'_{\rm syn} = 13.3 \ {\rm days} \, \left(\frac{B}{0.082\,{\rm G}}\right)^{-2}
\left(\frac{x'_{\rm eq}}{10^5}\right)^{-1} \ .
\label{eq111}
\end{equation}
In order to relate the loss timescale to the observed variability timescale, we need to transform into the observer's frame using Equation~(\ref{eq84}), which yields
\begin{equation}
t_{\rm syn} = 6.4 \ {\rm hours} \,
\left(\frac{\delta_{\rm D}}{50}\right)^{-1}
\left(\frac{B}{0.082\,{\rm G}}\right)^{-2}
\left(\frac{x'_{\rm eq}}{10^5}\right)^{-1} (1+z) \ .
\label{eq112}
\end{equation}
Hence we conclude that in the observer's frame, the synchrotron variability timescale is comparable to the variability timescale for the flare.

We can also perform a similar calculation based on the MHD acceleration timescale, given in the co-moving frame by
\begin{equation}
t'_{_{\rm MHD}} = \frac{x'}{<\!dx'/dt'\!>_{_{\rm MHD}}} \ , \qquad
\left<\!\frac{dx'}{dt'}\!\right>_{_{\rm MHD}} = 4 D_0 x' \ ,
\label{eq113}
\end{equation}
where the final result follows from Equation~(\ref{eq74}). Transforming into the observer's frame yields
\begin{equation}
t_{_{\rm MHD}} =  1.4 \ {\rm hours} \,
\left(\frac{\delta_{\rm D}}{50}\right)^{-1}
\left(\frac{D_0}{10^{-6}\,{\rm s}^{-1}}\right)^{-1} (1+z) \ .
\label{eq114}
\end{equation}
Equations~(\ref{eq112}) and (\ref{eq114}) imply that the electrons are able to achieve an approximate equilibrium distribution during the flare, and therefore it is reasonable to model the peak flare spectrum using a steady-state calculation such as the one developed in Section~4.

It is important to emphasize that the conclusions reached here only apply to the {\it continually} injected electrons. On the other hand, the {\it impulsively} injected electrons associated with the time lags will not achieve an equilibrium distribution, because the value of $D_0$ is smaller, and also because there is no continual particle injection to balance losses and escape. We plan to explore all of these issues in future work using a fully time-dependent simulation.

\subsection{Magnetization Parameter}

The level of stochastic acceleration experienced by the electrons due to collisions with MHD waves is regulated by the value of the momentum-diffusion coefficient, $D_0$, which is determined as part of  our qualitative fitting approach. Separate values for $D_0$ are obtained from the analysis of the time lag data and the spectral data, as discussed in Sections~6.1 and 6.2. The quantity $D_0$ is related to the MHD coherence length, $\ell_{_{\rm MHD}}$, via Equation~(\ref{eq17}), which can be rewritten as
\begin{equation}
\ell_{_{\rm MHD}}=\frac{c \, \sigma_{\rm mag}}{3D_0} \ ,
\label{eq115}
\end{equation}
where the magnetization parameter, $\sigma_{\rm mag}$, is defined by (Cerutti et al. 2012; Sironi et al. 2013; Sironi \& Spitkovsky 2014)
\begin{equation}
\sigma_{\rm mag} \equiv \left( \frac{v_{\rm A}}{c} \right)^2 \ .
\label{eq116}
\end{equation}
Our treatment of spatial diffusion in the field of MHD waves is valid provided the coherence length $\ell_{_{\rm MHD}}$ is smaller than the size of the plasma blob, so that
\begin{equation}
\ell_{_{\rm MHD}} \le R' \ ,
\label{eq117}
\end{equation}
which can be combined with Equations~(\ref{eq115}) and (\ref{eq116}) to derive a constraint on the magnetization parameter constraint, given by
\begin{equation}
\sigma_{\rm mag} \le \sigma_{\rm max} \equiv \frac{3 D_0 R'}{c} \ .
\label{eq118}
\end{equation}
The results obtained for $\sigma_{\rm max}$ are reported in Table~1. We generally find that $\sigma_{\rm max} \sim 0.1$, which is consistent with the values of $\sigma_{\rm mag}$ deduced observationally by Zhang et al. (2013), and theoretically by Zdziarski et al. (2015).

\subsection{Fermi Acceleration vs. Adiabatic Losses}

The mean Fermi particle acceleration rate in the co-moving frame of the outflowing plasma blob is obtained by setting $b=0$ in Equation~(\ref{eq74}), which yields
\begin{equation}
\left< \frac{d x'}{d t'} \right>_{\rm F} = \left< \frac{d x'}{d t'} \right>_{\rm F1}
+ \left< \frac{d x'}{d t'} \right>_{\rm F2} \ ,
\label{eq119}
\end{equation}
where
\begin{equation}
\left< \frac{d x'}{d t'} \right>_{\rm F1} = a D_0 \, x' \ , \qquad
\left< \frac{d x'}{d t'} \right>_{\rm F2} = 4 D_0 \, x' \ ,
\label{eq120}
\end{equation}
denote the mean first- and second-order Fermi acceleration rates, respectively. According to Equation~(\ref{eq119}), the dimensionless theory parameter $a$ represents the total first-order Fermi acceleration rate due to shock acceleration and adiabatic losses experienced by the electrons in the expanding jet outflow. We can express the total first-order Fermi parameter $a$ as the sum of two components, $a_{\rm sh}$ and $a_{\rm ad}$, corresponding to shock acceleration and adiabatic losses, respectively, by writing
\begin{equation}
a = a_{\rm sh} + a_{\rm ad} \ ,
\label{eq121}
\end{equation}
where (see Equations~(\ref{eq24}), (\ref{eq25}), (\ref{eq26}), and (\ref{eq40}))
\begin{equation}
a_{\rm sh} \equiv \frac{A^{\rm sh}_0}{D_0}
= \frac{u_-^2}{c\,\chi\ell_{_{\rm MHD}} D_0}\left( \frac{\chi - 1}{\chi + 1} \right)
\ ,
\label{eq122}
\end{equation}
and
\begin{equation}
a_{\rm ad} \equiv \frac{A^{\rm ad}_0}{D_0}
= - \frac{1}{3 D_0} \, (\vec{\nabla} \cdot \vec v)
\ .
\label{eq123}
\end{equation}
We discuss each of these processes separately below.

\subsubsection{Adiabatic Losses}

Adiabatic losses occur as the result of the expansion of the plasma blob as it propagates outward through the jet. The mean rate of change of the electron momentum due to adiabatic losses is given by Equation~(\ref{eq25}), which can be rewritten in the co-moving frame notation as
\begin{equation}
\left<\!\frac{dp'}{dt'}\!\right>\bigg|_{\rm ad} = - \frac{1}{3V'} \frac{dV'}{dt'} \, p'
= - \frac{1}{R'} \frac{dR'}{dt'} \, p' \ ,
\label{eq124}
\end{equation}
where $V'$ and $R'$ denote the co-moving volume and radius of the blob, respectively, and $V' \propto R'^3$. We can estimate the rate of expansion of the blob by making the generic assumption that the jet is conical and expanding with a constant velocity in the emission region. In this case, the co-moving radius of the blob, $R'$, should scale with the co-moving time, $t'$, and therefore Equation~(\ref{eq124}) reduces to
\begin{equation}
\left<\!\frac{dp'}{dt'}\!\right>\bigg|_{\rm ad} = - \frac{d\ln R'}{d\ln t'} \, \frac{p'}{t'_{\rm elap}}
= - \frac{p'}{t'_{\rm elap}} \ .
\label{eq125}
\end{equation}
Here, $t'_{\rm elap}$ denotes the elapsed time in the co-moving frame, which is related to the elapsed time in the observer frame, $t_{\rm elap}$, via the special relativistic proper time transformation, combined with the cosmological redshift, which give $t_{\rm elap}=t'_{\rm elap}\Gamma (1+z)$, were $\Gamma$ is the bulk Lorentz factor for the jet. For highly relativistic jets, $\Gamma \gg 1$, and the half-angle $\theta \sim 1/\Gamma$, in which case one can show that $\Gamma\sim \delta_{\rm D}$ (e.g., Abdo et al. 2011).
Assuming propagation at essentially the speed of light over the length of the jet, we obtain
\begin{equation}
t_{\rm elap} = \frac{D_{\rm emiss}}{c} \ ,
\label{eq126}
\end{equation}
where $D_{\rm emiss}$ is the distance between the black hole and the emission region.

Combining Equations~(\ref{eq26}) and (\ref{eq125}), we can express the adiabatic momentum loss rate in the co-moving frame as
\begin{equation}
<\!\dot p'\!>_{\rm ad} = A^{\rm ad}_0 \, p' \ ,
\label{eq127}
\end{equation}
where the quantity $A^{\rm ad}_0$ is defined by 
\begin{equation}
A^{\rm ad}_0 \equiv - \frac{1}{t'_{\rm elap}}
= - \frac{\Gamma(1+z)}{t_{\rm elap}} \ \propto \ \rm s^{-1} \ .
\label{eq128}
\end{equation}
Since the timescale $t_{\rm elap}$ is on the order of a year, it is far larger than the $\sim 1\,$hour variability timescales of interest here, and therefore we can safely treat $A^{\rm ad}_0$ as a constant during the X-ray flares from Mrk~421.

We can now combine Equations~(\ref{eq123}), (\ref{eq126}), and (\ref{eq128}) to obtain an expression for the dimensionless adiabatic loss parameter, $a_{\rm ad}$, given by
\begin{equation}
a_{\rm ad} = \frac{A^{\rm ad}_0}{D_0} = - \frac{\Gamma (1+z) c}{D_0 \, D_{\rm emiss}} \ ,
\label{eq129}
\end{equation}
or, equivalently,
\begin{equation}
a_{\rm ad} = - 4.9 \left(\frac{D_0}{10^{-6}\,{\rm s}^{-1}}\right)^{-1}
\left(\frac{\Gamma}{50}\right) \left(\frac{D_{\rm emiss}}{0.1\,{\rm pc}}\right)^{-1}
(1+z) \ .
\label{eq130}
\end{equation}
As discussed in Section~1, we expect that the emission distance $D_{\rm emiss}$ falls in the range $0.1\,{\rm pc} \lesssim D_{\rm emiss} \lesssim 1\,$pc. Setting $D_0 = 8.49 \times 10^{-7}\,{\rm s}^{-1}$ for the peak flare spectrum (see Table~1), $z=0.031$, and $\Gamma\sim\delta_{\rm D}=50$, we find that the corresponding range of values for $a_{\rm ad}$ obtained using Equation~(\ref{eq130}) is $-5.9 \lesssim a_{\rm ad} \lesssim -0.59$. This range includes the value for the total first-order Fermi parameter, $a = -3.3$, obtained by fitting our model to the peak flare X-ray spectrum in Section~6.2. Since $a = a_{\rm sh} + a_{\rm ad}$ (see Equation~(\ref{eq121})), we conclude that during the formation of the peak flare spectrum, adiabatic losses dominate over shock acceleration for the electrons continually injected throughout the blob. However, it is important to note that due to the additional particle acceleration provided via stochastic wave-particle interactions, with mean acceleration rate (see Equation~(\ref{eq119})) $<\!dx'/dt'\!>_{\rm F2} = 4D_0 x'$, the total Fermi acceleration rate given by Equation~(\ref{eq119}) is still positive, with the value $<\!dx'/dt'\!>_{\rm F} = (4-3.3)D_0 x'=0.7 D_0 x'$. This implies that second-order (stochastic) particle acceleration experienced by the at-large electrons distributed throughout the blob powers the production of the peak flare X-ray spectrum, rather than shock acceleration.

\subsubsection{Shock Acceleration}

The value of the theory parameter $a$ obtained in the steady-state spectrum calculation discussed in Section~7.3.1 is $a=-3.3$, which is consistent with adiabatic cooling in the jet, and suggests that shock acceleration is unimportant in the formation of the peak flare X-ray spectrum. We obtain a completely different result in the calculation of the time lags discussed in Section~6.1, where we show that $a = 40.0$. This large positive value for $a$ indicates that first-order Fermi acceleration at the shock overwhelms adiabatic losses for the impulsively-injected electrons that generate the observed X-ray time lags from Mrk~421. The adiabatic loss rate should be the same for both the steady-state and transient electron populations, and therefore we set $a_{\rm ad}$=-3.3 for the time lag calculation. Since $a = a_{\rm sh} + a_{\rm ad}$ according to Equation~(\ref{eq121}), this implies that the value of the shock acceleration parameter is $a_{\rm sh}=43.3$ during the formation of the transient electron population that produces the time lags. Let us explore whether this value for $a_{\rm sh}$ is consistent with the physics of the shock propagating through the plasma blob during the formation of the time lags.

Using Equation~(\ref{eq17}) to eliminate the product $\ell_{_{\rm MHD}} D_0$ in Equation~(\ref{eq122}) yields the alternative form
\begin{equation}
a_{\rm sh} \equiv \frac{A^{\rm sh}_0}{D_0}
= \frac{3}{\chi} \left( \frac{\chi - 1}{\chi + 1} \right)
\left(\frac{u_-}{c}\right)^2 \left(\frac{v_{\rm A}}{c}\right)^{-2}
\ ,
\label{eq131}
\end{equation}
where $\chi$ is the shock compression ratio and $u_-$ is the upstream flow velocity in the frame of the shock. Setting $\chi=4$ for a strong shock, we obtain an estimate of the shock acceleration parameter $a_{\rm sh}$, given by
\begin{equation}
a_{\rm sh} \sim 45  \left(\frac{u_-}{c}\right)^2 \left(\frac{\sigma_{\rm mag}}{0.01}\right)^{-1}
\ ,
\label{eq132}
\end{equation}
where the magnetization parameter $\sigma_{\rm mag}$ is defined in Equation~(\ref{eq116}). Setting $u_- \sim c$ for a mildy relativistic shock, and adopting the observational the estimate $\sigma_{\rm mag} \sim 0.01$ for Mrk~421 from Zhang et al. (2013), we obtain the estimate $a_{\rm sh} \sim 45$. This value agrees remarkably well with the result $a_{\rm sh} = 43.3$ obtained by fitting our theoretical model to the time lag data obtained during the 1998 April 21 flare from Mrk~421.

\subsection{Maximum Larmor Radius}

In Section 4.2, we derived the Fokker-Planck form of the steady-state transport equation, with broadening and drift coefficients given by Equations (\ref{eq73}) and (\ref{eq74}), respectively. The associated value for the equilibrium Lorentz factor, $x'_{\rm eq}$, is given by Equation~(\ref{eq86}), and reported in Table~1. We generally find that $x'_{\rm eq} \sim 10^5$, which implies that low-energy electrons will be accelerated up to this characteristic energy before synchrotron losses become important, leading to the exponential turnover seen in the electron distribution in Figure~3. We can compute the maximum Larmor radius, $r_{\rm L}^{\rm max}$, corresponding to the equilibrium Lorentz factor $x'_{\rm eq}$, by using Equation (\ref{eq5}) to write
\begin{equation}
r_{\rm L}^{\rm max} \equiv \frac{x'_{\rm eq}m_ec^2}{q B}
= 2.08 \times 10^9 \, {\rm cm} \left(\frac{x'_{\rm eq}}{10^5} \right) \left(\frac{B}{0.082} \right)^{-1} \ .
\label{eq133}
\end{equation}
Based on this relation, we conclude that $r_{\rm L}^{\rm max} \ll R'$, where $R' = 5.3 \times 10^{15}\,$cm is the radius of the blob, for both the time-lag and peak-spectrum calculations. This condition, when combined with Equation~(\ref{eq7}), also ensures that the diffusion velocity, $w_{\rm L}$, is far below the speed of light, as required, and therefore our utilization of the diffusion approximation is justified.

\section{DISCUSSION AND CONCLUSION}

The observation of X-ray flares from blazars raises interesting theoretical questions regarding the nature of the particle acceleration mechanism and the ultimate power source for the flares. Previous attempts to interpret the data and deduce the nature of the underlying electron population by ``reverse-engineering'' the X-ray spectrum have led to an approximate determination of the shape of the electron distribution, along with estimates for other source parameters, such as the co-moving blob radius $R'$, the magnetic field strength $B$, and the Doppler factor $\delta_{\rm D}$ (e.g., Fossati et al. 2000b; Abdo et al. 2011; Finke et al. 2008). The theoretical picture has been further challenged by observations of X-ray time lags during some flares, such as the 1998 April 21 flare from Mrk~421 studied by Zhang (2002).

The combination of the spectral data with the time lags comprise a set of observations that are very difficult to understand in the absence of a detailed physical model that includes time-dependent particle acceleration. In this paper, we have developed a new analytical, first-principles physical model describing the transport and acceleration of relativistic electrons injected into a blob of plasma propagating outward through a blazar jet. Our goal in this work is to use a single integrated model to simultaneously explain the formation of the X-ray time lags observed from Mrk~421 during the 1998 April 21 flare, as well as the X-ray spectrum observed at the peak of the flare.

\subsection{Integrated Model for Time Lags and X-ray Spectrum}

The model developed here envisions an outflowing plasma blob with co-moving radius $R'$ containing radiating plasma, magnetic fields, and shocks, moving towards the observer with Doppler factor $\delta_{\rm D} \gg 1$. The model is represented by a transport equation that includes terms describing first-order Fermi acceleration due to shocks, second-order (stochastic) Fermi acceleration due to MHD wave-particle interactions, losses due to adiabatic expansion, losses due to synchrotron emission, and particle escape regulated by Bohm diffusion. By averaging over the volume of the blob, we developed a simplified, one-zone spatial model that is similar to those employed in a number of previous studies (e.g., Finke et al. 2008). The transport equation was solved in Sections~3 and 4, respectively, to obtain exact solutions corresponding to the time-dependent and steady-state cases. The exact solutions were used to model the time lags in Section~6.1, and the peak flare X-ray spectrum in Section~6.2. In both calculations, we used the same value for the blob radius, $R' = 5.3 \times 10^{15}\,$cm, the Doppler factor, $\delta_{\rm D} = 50$, and the magnetic field, $B = 0.082\,$G.

The model parameters for the two calculations are listed in Table~1. We find that the time lags require a very large value for the first-order Fermi acceleration parameter, $a=40.0$, whereas the spectrum calculation requires a negative value for $a$, given by $a=-3.3$ (see Section~6). The physical interpretation of these two very different values for $a$ was discussed in detail in Section~7, where we concluded that the transient time lag signal is generated by electrons strongly accelerated in the vicinity of a shock wave (or waves) inside the blob. On the other hand, the peak X-ray spectrum is generated by electrons that are continually injected throughout the blob, and these electrons primarily experience a combination of stochastic MHD wave-driven acceleration, as well as synchrotron losses.

Another important difference between the two calculations is that the Lorentz factor of the injected seed electrons is $x'_0 = 2.55 \times 10^5$ in the time-lag calculation, versus $x'_0=2$ in the spectrum calculation. This clearly suggests two very different origins for the two populations of seed electron. It is interesting to note that the values of the momentum diffusion coefficient $D_0$ obtained in the two cases are similar, with $D_0 \sim 10^{-7}\,{\rm s}^{-1}$, implying that the two populations of electrons see similar distributions of randomly propagating MHD waves. However, the different values for $a$ in the two scenarios imply that the two populations of electrons experience very different levels of first-order Fermi acceleration and adiabatic losses.

We propose that the results described above can naturally be explained in terms of a two-component model for the electron population as follows. The dominant component of the electron distribution generates $\sim 90\%$ of the X-ray signal observed during the flare, and varies on relatively long timescales of $\sim 1\,$day. We propose that this component of the electron distribution represents particles continually injected throughout the blob. These particles are picked up from the high-energy tail of the thermal electron distribution that permeates the entire blob, with initial Lorentz factor $x'_0=2$. After injection, they are accelerated mainly via interactions with a random field of MHD waves propagating along the local magnetic field, leading to strong second-order (stochastic) Fermi acceleration. On average, these at-large electrons do not interact very frequently with the shock waves in the blob, but they do experience stochastic acceleration, in addition to synchrotron losses, and adiabatic losses associated with the expansion of the blob in the jet outflow.

The second component of the electron distribution is the population injected at the shock wave, with a very high initial Lorentz factors, $x'_0= 2.55 \times 10^5$. We hypothesize that these high-energy seed electrons are generated as a result of magnetic reconnection occurring in the vicinity of the shock. After injection they experience adiabatic losses, stochastic acceleration, and synchrotron/inverse Compton losses, but the strong shock acceleration is able to raise their Lorentz factors up to $x'_{\rm eq} \sim 10^6$. During the transient acceleration phase, hard time lags develop in this electron population since it is unable to achieve equilibrium. The hard lags in the electron distribution become imprinted on the photon distribution via the emission of synchrotron radiation at higher and higher energies during the transient. Since the second component produces $\sim 10\%$ of the observed X-ray emission (with a variability timescale of $\sim 1\,$hour), we assume that the number of impulsively-injected seed electrons is $\sim 10\%$ of the total number of injected electrons.

\subsection{Physical Interpretation of Time Lags}

The appearance of the sharp transition to a soft time lag above the Fourier frequency $\nu_f \sim 10^{-4.05}\,$Hz plotted in Figure~1 warrants further discussion. In our model, the transient electrons are injected with Lorentz factor $x'_0 = 2.55 \times 10^5$ as measured in the co-moving frame of the blob. At the instant of injection, the electrons emit synchrotron photons with energy $\epsilon$ measured in the observer's frame, given by (see Equation~(\ref{eq85}))
\begin{equation}
\epsilon = {x'_0}^2 \, m_e c^2 \, \frac{B \xi}{B_c} \frac{\delta_{\rm D}}{1+z}
\ .
\label{eq85new}
\end{equation}
Setting $x'_0 = 2.55 \times 10^5$, $\xi=1$, $B=0.082$, $z=0.031$, and $\delta_{\rm D}=50$ yields $\epsilon=3.19\,$keV in the observer's frame. This energy falls within the high-energy window utilized by Zhang (2002), which extends from 2.0-10.0\,keV, and it  is outside the low-energy window, which extends from 0.1-2.0\,keV. Hence, at first glance, it seems surprising that our computational results predict a hard lag rather than a soft lag, since the initial instantaneous emission is detected in the observer's high-energy window. In order to better understand this apparent paradox, it is instructive to consider the time series of the data detected by the observer at the channel energies $\epsilon_s=1.05\,$keV and $\epsilon_h=6.00\,$keV that we use in our computations.

The closed-form solution for the Fourier transform of the observed flux, $\tilde{\mathscr{F}}(\epsilon,\omega)$, is given by Equation~(\ref{eq103}), and the actual observed flux, ${\mathscr{F}}(\epsilon,t)$, is therefore given by the inverse Fourier transform (see Equations~(\ref{eq42}) and (\ref{eq96}))
\begin{equation}
{\mathscr{F}}(\epsilon,t) = \frac{1}{2\pi}\int_{-\infty}^{\infty} e^{-i \omega t} \tilde{\mathscr{F}}(\epsilon,\omega)
d\omega \ .
\label{eq42new}
\end{equation}
In Figure~4, we plot the results obtained for the observed flux (normalized to a peak value of unity) using Equation~(\ref{eq42new}) in the energy channels $\epsilon_s=1.05\,$keV and $\epsilon_h=6.00\,$keV as a function of the elapsed time $t$ since the impulsive particle injection, using the same theoretical parameters that were used to compute the time lags presented in Figure~1. Note that at small times, the evolution is dominated by a fast rise, {\it which occurs in the hard channel first}. This rapid variability is associated with the highest Fourier frequencies in Figure~1, and this explains the soft (negative) time lags above $\nu_f\sim10^{-4.05}\,$Hz. On the other hand, at later times, the evolution of the light curves is dominated by a gradual exponential decline {\it which occurs in the soft channel first}. This slow evolution corresponds to low Fourier frequencies, and this explains the hard times lags observed in Figure~1 for $\nu_f \lapprox 10^{-4.05}\,$Hz. The sharp transition from hard to soft lags at frequency $\nu_f \sim 10^{-4.05}\,$Hz cannot be computed precisely because the Fourier time lag is the result of a nonlinear calculation. However, we note that this frequency value is reasonable, since one would naively expect that the critical frequency would be comparable to 1/(5000 sec), corresponding to the approximate midpoint of the light curves, where the maxima occur. Hence we conclude that the light curves plotted in Figure~4 fully support the time lag results presented in Figure 1, and in particular, they are consistent with the hard time lags we obtain at low Fourier frequencies.

Another interesting question is whether the finite light-travel time across the blob, $R'/c$, could introduce modifications in the Fourier time lag profiles that have not been considered in our analysis. In the one-zone model assumed here, the finite light-travel time would essentially cause a time-dependent radiation wave to appear to move across the surface of the blob, as seen by a distant observer. The question is whether this phenomenon would create any observable changes in the time lags we compute here. The answer is that it would not, because the propagation of the radiation wave is a coherent phenomenon between the energy channels. This means that the resulting additional Fourier phase lag introduced by the light-crossing time effect is independent of the observing energy. Since the phase lag is energy independent, it follows that there is no additional time lag introduced between the two energy channels by the light-travel time effect. This can be seen mathematically by looking at Equation (47) from Finke \& Becker (2014). The light-travel time effect is represented by the leading exponential factor, which has a complex exponent. However, the exponent is not a function of photon energy. Hence, when one constructs the complex cross spectrumß between the hard and soft channels, there is no energy-dependent phase shift, and therefore no time lag, related to the light-crossing time.

\subsection{Variation of Channel Energies}

Since Zhang (2002) utilized continuous energy windows for his computations of the time lags, rather than precise values. Our primary results for the time lags, plotted in Figure~1, were obtained by setting the soft and hard channel energies in our model equal to Zhang's channel-center energies, so that $\epsilon_s=1.05\,$keV and $\epsilon_h=6.00\,$keV. Since this choice of energies is somewhat arbitrary, it is interesting to examine the effect of utilizing alternative values for $\epsilon_s$ and $\epsilon_h$. Here, we present two such alternative models, corresponding to $\epsilon_s=0.90\,$keV and $\epsilon_h=6.47\,$keV, and $\epsilon_s=1.20\,$keV and $\epsilon_h=5.60\,$keV, respectively. The resulting time lag profiles are plotted in Figure~5, and compared with the ``primary'' time lag profile, which is the result plotted in Figure~1, obtained by setting $\epsilon_s=1.05\,$keV and $\epsilon_h=6.00\,$keV. All three time lag calculations utilized the same theoretical parameter values, listed in Table~1. We can see that the qualitative fits to the data are acceptable in all three cases.

\subsection{Conclusion}

The relatively simple model developed here can successfully account for both the formation of the peak flare X-ray spectrum, and the X-ray time lags, for the 1998 April 21 flare from Mrk~421. The theoretical parameter values implied by our model, and reported in Table~1, are very close to those obtained using independent observational estimates in Section~7. We plan to further refine the model in future work, including the incorporation of a more accurate calculation of the synchrotron spectrum using the exact integral, in place of the $\delta$-function approximation employed here. Additional modifications include a complete implementation of the nonlinear SSC calculation, which was neglected here since our model is based on a linear transport equation. We also anticipate the possibility of developing a fully time-dependent calculation that would yield result for the time-variable X-ray spectrum, which would further reinforce and extend the results presented here. We plan to pursue these modifications in future work.

The authors are grateful to the anonymous referee for making several insightful observations that stimulated significant improvements in the manuscript, especially regarding the interpretation of the time lag results, and the associated light curves. J. D. F. acknowledges support from the Chief of Naval Research.

\begin{deluxetable}{cccccccccccr}
\tabletypesize{\scriptsize}
\tablecaption{Model Parameters}
\label{table1}
\tablewidth{0pt}
\tablehead{
\colhead{Variable}
& \colhead{Time Lag Model}
& \colhead{Flare Spectrum Model}
}

\startdata
$z$
&$0.031$
&$0.031$
\\
$B\,$(G)
&$0.082$
&$0.082$
\\
$R\,$(cm)
&$5.30 \times10^{15}$
&$5.30 \times10^{15}$
\\
$\delta_{\rm D}$
&50.0
&50.0
\\
\\
$x'_0$
&$2.55 \times 10^5$
&$2$
\\
$a$
&$40.0$
&$-3.30$
\\
$b$
&$7.94 \times 10^{-5}$
&$1.02 \times 10^{-5}$
\\
$\tau$
&$4.93 \times 10^9$
&$3.83 \times 10^{10}$
\\
$\dot N _0\,({\rm s}^{-1})$
& N/A
&$2.82 \times 10^{34}$
\\
\\
$A_0\,({\rm s}^{-1})$
&$4.38 \times 10^{-6}$
&$-2.80 \times 10^{-6}$
\\
$B_0\,({\rm s}^{-1})$
&$8.69 \times 10^{-12}$
&$8.69 \times 10^{-12}$
\\
$D_0\,({\rm s}^{-1})$
&$1.09 \times 10^{-7}$
&$8.49 \times 10 ^{-7}$
\\
$\sigma_{\rm max}$
&0.058
&0.45
\\
$x'_{\rm eq}$
&$5.54 \times 10^5$
&$6.84 \times 10^4$
\\
\enddata
\end{deluxetable}

\begin{figure}[ht]
\vspace{0.0cm}
\centering
\includegraphics[width=0.5\textwidth]{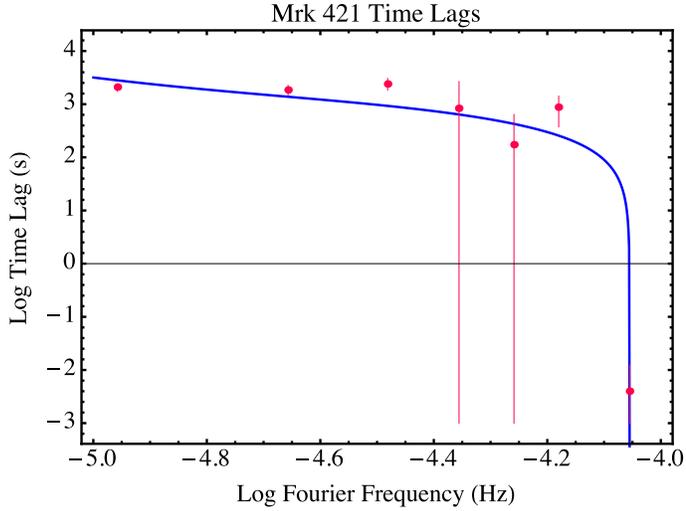}
\caption{Theoretical X-ray time lag profiles, $\delta t$, plotted as a function of the Fourier frequency $\nu_f=\omega/(2 \pi)$ using Equations~(\ref{eq1}) and (\ref{eq4}), with the hard and soft Fourier components, $G_h$ and $G_s$, respectively, evaluated using Equations~(\ref{eq105}). A positive lag indicates that the hard X-ray signal is delayed relative to the soft signal. The hard and soft channel energies are $\epsilon_h=6\,$keV and $\epsilon_s=1.05\,$keV, respectively. Also plotted for comparison are the time lags computed by Zhang (2002) using {\it Beppo}SAX data for the 1998 April 21 flare of Mrk~421, with 1$\sigma$ error bars. The corresponding theory parameters are listed in Table~1.}
\end{figure}

\begin{figure}[ht]
\vspace{0.0cm}
\centering
\includegraphics[width=0.5\textwidth]{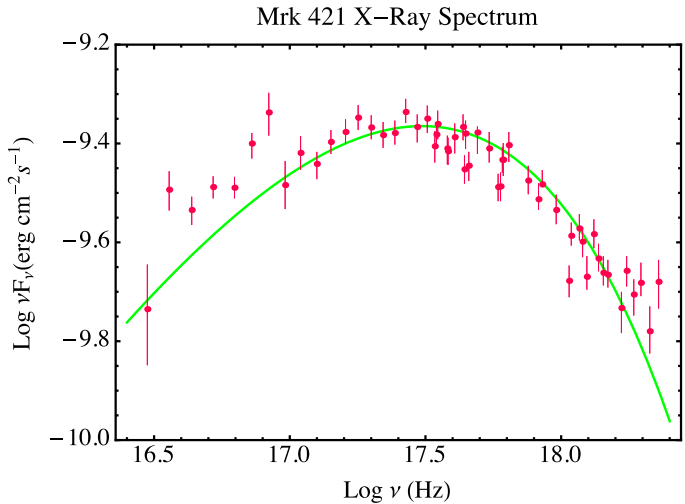}
\caption{Steady-state X-ray flux function $\mathscr{F} \equiv \nu F_\nu$, plotted as a function of the photon frequency $\nu$, evaluated using Equation~(\ref{eq95}). The associated theory parameters are listed in Table~1. Also plotted is the observed X-ray spectrum for the 1998 April 21 flare, taken from Figure~3 in Fossati et al. (2000b). The theoretical X-ray spectrum agrees fairly well with the observational data.}
\end{figure}

\begin{figure}[ht]
\vspace{0.0cm}
\centering
\includegraphics[width=0.5\textwidth]{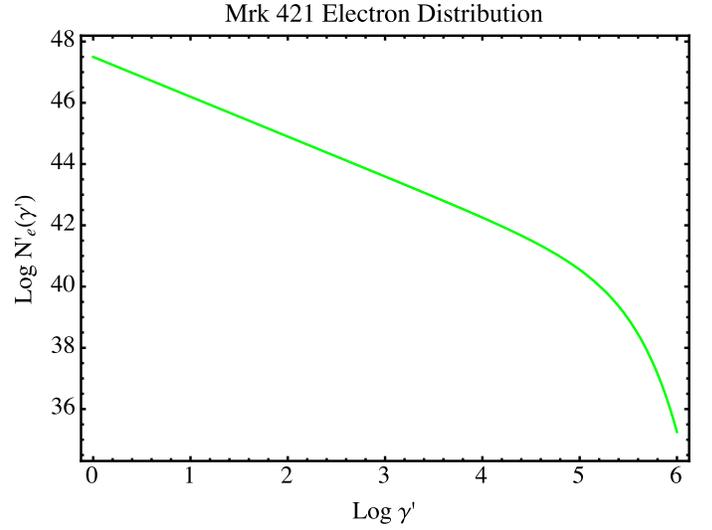}
\caption{Steady-state electron number distribution, $N_e'(\gamma')$, plotted as a function of the Lorentz factor $\gamma'$, as seen in the co-moving frame of the plasma blob, evaluated using Equation~(\ref{eq92}). The model parameters are the same as those used to compute the X-ray spectrum in Figure~2, and are listed in Table~1. The number distribution has a power-law shape up to the exponential cutoff at the equilibrium Lorentz factor, $x'_{\rm eq} = 6.84 \times 10^4$.}
\end{figure}

\begin{figure}[ht]
\vspace{0.0cm}
\centering
\includegraphics[width=0.5\textwidth]{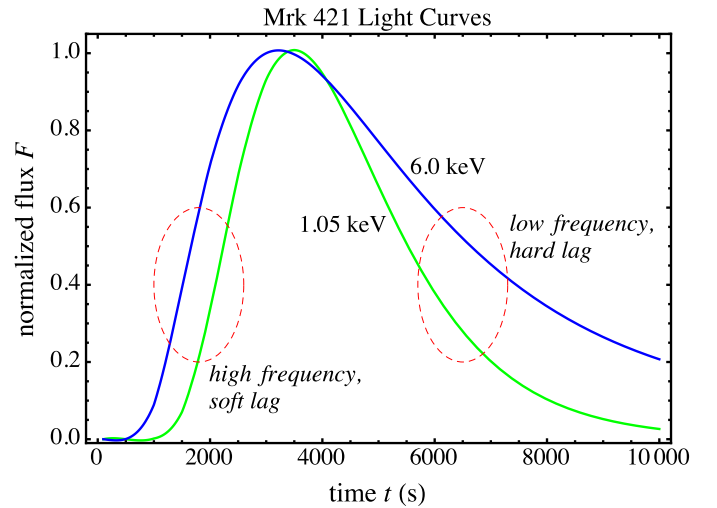}
\caption{Light curves in the soft and hard energy channels, $\epsilon_s=1.05\,$keV and $\epsilon_h=6.00\,$keV, respectively, plotted as a function of the elapsed time $t$ in seconds since injection. The curves were computed using Equation~(\ref{eq42new}), based on the same theoretical parameters used to generate the time lags in Figure~1 (see Table~1). The fast initial rise of the light curves occurs in the {\it hard channel first}, and therefore we would expect to observe a {\it soft time lag} at high Fourier frequencies, in agreement with Figure~1. Conversely, the gradual exponential decline of the light curves at later times occurs in the {\it soft channel first}, and this explains the {\it hard time lags} observed in Figure~1 at low Fourier frequencies.}
\end{figure}

\begin{figure}[ht]
\vspace{0.0cm}
\centering
\includegraphics[width=0.5\textwidth]{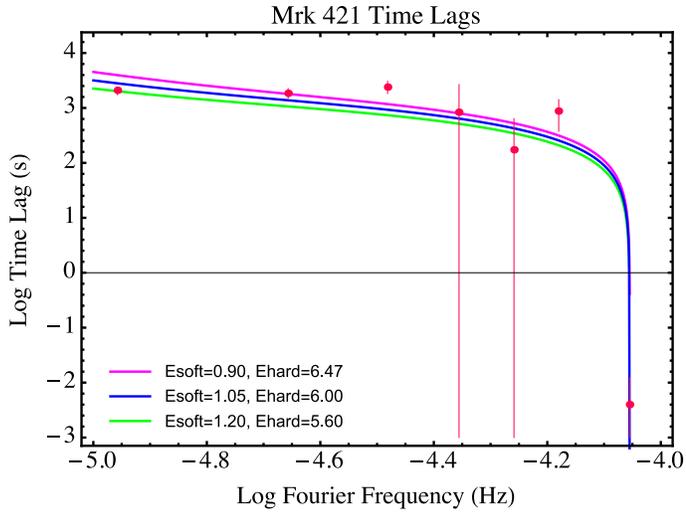}
\caption{Same as Figure~1, except we have used three different sets of values for the hard and soft channel energies, $\epsilon_h$ and $\epsilon_s$, respectively, as indicated in keV for each curve. All three profiles were computed using the same theory parameters, listed in Table~1, and the blue curve is the same result plotted in Figure~1.}
\end{figure}

{}

\label{lastpage}

\end{document}